\documentclass[manuscript]{emulateapj}

\usepackage{natbib}
\usepackage{lscape}
\usepackage{rotating}
\bibpunct{(}{)}{,}{a}{}{,}

\slugcomment{ApJ accepted}

\shorttitle{\emph{K2}: A new cluster detector}
\shortauthors{Thanjavur, Willis \& Crampton}

\begin{document}
\def\asec{^{\prime\prime}}
\def\mag{\:{\mathrm mag}}
\def\angs{\mathrm{\AA}}
\def\sqdeg{\:\mathrm{deg^2}}
\def\deg{\ensuremath{^\circ}}
\def\omegam{$\Omega_{\mathrm m}$}
\def\omegab{$\Omega_{\mathrm b}$}
\def\omegabh{$\Omega_{\mathrm b} h^2$}
\def\omegal{$\Omega_{\Lambda}$}
\def\ovrdn{$\Delta_{200}$}
\def\ovrdnr{$\Delta(r)$}
\def\ovrdnv{$\Delta_{vir}$}
\def\vcr{$v_{\mathrm c}(r)$}
\def\vc{$v_{\mathrm c}(r)$}
\def\halpha{$\mathrm H_{\alpha}$}
\def\rhoc{$\rho_{\mathrm crit}$}
\def\rhocore{$\rho_{0}$}
\def\rcore{$r_{\mathrm c}$}
\def\rhobar{$\bar{\rho}$}
\def\rhor{$\rho(r)$}
\def\mr{$M(r)$}
\def\radius{$r$}
\def\rscale{$r_{\mathrm s}$}
\def\deltac{$\delta_{\mathrm c}$}
\def\lcdm{$\Lambda$CDM }
\def\scdm{$S$CDM }
\def\rnfw{$r^{-1}$}
\def\mathrmoore{$r^{-1.5}$}
\def\hub{$h$}
\newcommand{\mc}[3]{\multicolumn{#1}{#2}{#3}}
\def\G{{\mathrm G}}
\def\c{{\mathrm c}}
\def\gsim{ \lower .75ex \hbox{$\sim$} \llap{\raise .27ex \hbox{$>$}} }
\def\lsim{ \lower .75ex \hbox{$\sim$} \llap{\raise .27ex \hbox{$<$}} }
\def\Mpc{{\mathrm Mpc}}
\def\kpc{{\mathrm kpc}}
\def\pc{{\mathrm pc}}
\def\Lsun{{\mathrm L_\odot}}
\def\Msun{\:{\rm M_\odot}}
\def\gsim{\ga}
\def\eg{e.g.\ }
\def\lsim{\la}
\def\etal{{et al.\ }}
\def\mpc {\mathrm{Mpc}}
\def\kpc {\mathrm{kpc}}
\def\msun {\:{\rm M_\odot}}
\def\ergs {{\mathrm erg} \, {\mathrm s}^{-1}}
\def\cm {{\mathrm cm}}
\def\kms {{\mathrm {\,km \, s^{-1}}}}
\def\Hz {{\mathrm Hz}}
\def\yr {{\mathrm yr}}
\def\gyr {{\mathrm Gyr}}
\def\arcmin {{\mathrm arcmin}}
\def\G {{ G  }}
\def\magasec {\mathrm {\, mag.arc.sec^{-2}}} 
\def\lsim{\mathrel{\hbox{\rlap{\hbox{\lower4pt\hbox{$\sim$}}}\hbox{$<$}}}}
\def\gsim{\mathrel{\hbox{\rlap{\hbox{\lower4pt\hbox{$\sim$}}}\hbox{$>$}}}}

\def\and   {\mathrm {et al.} \mathrm}  
\def\mathrmd {\mathrm d}



\title{\emph{K2}: A NEW METHOD FOR THE DETECTION OF GALAXY CLUSTERS BASED ON CFHTLS MULTICOLOR IMAGES}

\author{Karun Thanjavur\altaffilmark{1,2}, Jon Willis\altaffilmark{1},
	and David Crampton\altaffilmark{1,2}}

\altaffiltext{1}{Department of Physics \& Astronomy, University of Victoria, 
        Victoria, BC, V8P 1A1, Canada; karun@uvic.ca}
\altaffiltext{2}{National Research Council of Canada, Herzberg Institute of 
        Astrophysics, 5071 West Saanich Road, Victoria, BC, V9E 2E7, 
        Canada}

\begin{abstract}
 
We have developed a new method, \emph{K2}, optimized for the detection of galaxy clusters in multicolor images. Based on the Red Sequence approach, \emph{K2} detects clusters using simultaneous enhancements in both colors and position. The detection significance is robustly determined through extensive Monte-Carlo simulations and through comparison with available cluster catalogs based on two different optical methods, and also on X-ray data. \emph{K2} also provides quantitative estimates of the candidate clusters' richness and photometric redshifts. Initially \emph{K2} was applied to the two color ($gri$) 161 $\sqdeg$ images of the Canada-France-Hawaii Telescope Legacy Survey ÐWide  (CFHTLS-W) data.  Our simulations show that the false detection rate for these data, at our selected threshold, is only $\sim1\%$, and that the cluster catalogs are $\sim80\%$ complete up to a redshift of z = 0.6 for Fornax-like and richer clusters and to $z \sim0.3$ for poorer clusters.  Based on the $g, r$ and $i$-band photometric catalogs of the $Terapix\;T05$ release, 35 clusters/$\sqdeg$ are detected, with 1-2 Fornax-like or richer clusters every two square degrees. Catalogs containing data for 6144 galaxy clusters have been prepared, of which 239 are rich clusters. These clusters, especially the latter, are being searched for gravitational lenses -- one of our chief motivations for cluster detection in CFHTLS. The \emph{K2} method can be easily extended to use additional color information and thus improve  overall cluster detection to higher redshifts. The complete set of \emph{K2} cluster catalogs, along with the supplementary catalogs for the member galaxies, are available on request from the authors.

\end{abstract}


\section{Introduction}

\indent  Galaxy clusters are the most massive virialized structures in the present day universe, with masses ranging from a few times $ 10^{14}\msun$ for poor clusters to rich clusters at mass $\sim 10^{15}\Msun$ or higher. Occurring at the intersection of mass filaments, clusters continue to build up mass thorough accretion of infalling galaxies and inter-cluster gas. Since clusters occur at the high end of halo mass function, their observed mass function and its evolution with redshift have therefore found extensive use as test grounds for cosmogony and large scale structure formation models \citep{Vikh09, Mali08, Mant08, Pier03} and for constraining cosmological parameters \citep{Wik08, Glad07, Hark07, Sefu07}. In addition, clusters also provide excellent test beds for models describing environmental effects on galaxy evolution. Since the deep gravitational potential well of a cluster retains all the accreted baryonic matter, detailed observations of the member galaxy population and of the intra-cluster medium provide important clues toward the star formation history and the presence and strength of associated feedback processes on the evolution of galaxies in high density environments \citep{Naga07, Voit05}.

\indent Our focus on clusters is for yet another application - their high projected surface mass densities result in large gravitational lens cross sections \citep{Henn07, Smit01}, thus making them very efficient in lensing high redshift objects which lie in the direction of their lines of sight (e.g., \citet{Ebel09, Smai07}). Spectacular cluster lenses, such as Abell 2218 \citep{Pell92}, are well known examples of this phenomenon. The magnification boost offered by these cluster lenses, when combined with modern observational techniques offers a viable technique to resolve sub-galactic scales in the high redshift universe; for example, Gemini GMOS 2D integral field (IFU) spectroscopy \citep{Swin03, Swin06} and Keck LRIS longslit spectroscopy \citep{Pett00, Pett02} of lensed arcs have been used to map the kinematics, intensity of star formation, as well as the effect of feedback processes on the evolution of high redshift (z $\gg$ 1) galaxies. The number of such investigations is limited mainly by the lack of a significant sample of bright, lensed galaxies to carry out these studies. 

\indent In order to detect new cluster lenses, we have undertaken a search for clusters and groups of galaxies and any associated lenses in multi-color photometric data with large sky coverage. The search is carried out in two steps: first detecting galaxy clusters and groups, (which have a higher likelihood of lensing background objects due to their mass concentrations), and subsequently carrying out a dedicated search for strong lens images in these cluster regions. In this publication, planned as the first of two papers, we present details of our cluster detection methodology, results from the Monte Carlo calibrations for contamination and expected completeness of the search, and the cluster catalogs generated by the detector from $161 \sqdeg$ of imaging from the \textit{Canada France Hawaii Telescope Legacy Survey, Synoptic Wide} component (referred hence as, \textit{CFHTLS-W}). The lens search scheme and the resulting lens catalogs will be presented in a succeeding publication. 

\indent It must be added that the public release of data from several large recent surveys has consequently encouraged the rapid development of automated searches for strong lenses, ranging from \emph{quasars lenses} in Sloan Digital Sky Survey (SDSS) imaging \citep{Inad08}, to \emph{galaxy-galaxy scale lenses}, e.g., robotic searches in HST-ACS imaging archives \citep{Mars09, Bolt08}, in CFHTLS imaging \citep{Caba07}, and visual search in Cosmic Evolution Survey (COSMOS) fields  \citep{Faur08, Jack08}, and to \emph{group scale lenses}, e.g. in CFHTLS imaging \citep{Limo09} and SDSS imaging \citep{Kubo09}. The strong lensing candidates in these catalogs are fully complemented by the search we have undertaken for giant arcs in \emph{cluster scale lenses} in CFHTLS-Wide imaging. 

\indent The recently completed CFHTLS project is a collaboration between the CFHT corporation, le Centre National de la Recherche Scientifique de France, le Commissariat \`{a} l'Energie Atomique and the National Research Council of Canada. Organized with a PI-less open structure, this wide field \emph{imaging} survey comprises of three components: \emph{Deep Synoptic, Wide Synoptic} and the \emph{Very Wide} surveys, each with different scientific objectives and observing strategies. The principal goal of the \emph{Wide Synoptic} part, the data of which we use for the cluster search, is to map structure formation in the universe using cosmological weak lensing; a more complete description of the objectives of all three components as well as other pertinent information are provided at the CFHT website\footnote{http://www.cfht.hawaii.edu/Science/CFHLS/}. 

\indent The CFHTLS observations were carried out at the Canada-France-Hawaii 3.6m telescope, using the 1 square degree wide field imager, \emph{MegaPrime}. The CFHTLS-W aimed to map 195 square degrees in the five-filter \emph{ugriz} set\footnote{We do not use the `primes' in the notation of these MegaPrime $u^*g'r'i'z'$ filters, which closely match the Sloan \emph{u'g'r'i'z'} filters in their characteristics} and in four different sky patches (referred as W1, W2, W3 and W4) to a depth of $i \leq 24\,$mag. In addition, an image quality constraint of seeing better than $0\asec.9$ in r-filter was imposed so that the stacked images could be used to construct shear maps for weak lensing analysis, the principal science goal. For the CFHTLS-W, the survey strategy was to observe all the fields first with only the $gri$ filters, with follow-up observations carried out in $urz$. Table \ref{Tbl-1} is a summary of the completion statistics of the four CFHTLS-W fields at the end of the survey period in August 2008; the statistics listed in the table have been computed using the observational details provided at the \emph{Terapix}\footnote{http://terapix.iap.fr/} website for each field. The complete data processing, stacking of images and their astrometric and photometric calibrations are carried out at \emph{Terapix} and the complete set of data products, including the photometric catalogs which we use for our cluster detection, is archived and distributed through the \emph{Canadian Astronomy Data Centre\footnote{http://www1.cadc-ccda.hia-iha.nrc-cnrc.gc.ca/cadc/}} (CADC). Our principal objective is to implement a cluster detection method tuned for 2-color imaging data and with a well understood selection function, to build a large sample of clusters at a median redshift $z \sim 0.6$.  

 \indent To motivate the development of our cluster detector, the paper begins with a discussion on various published methods designed for \emph{optical} data in Section \ref{RevCS}. We then introduce our cluster detection method in Section \ref{ClusSearch}, highlighting its relevance and specific advantages for this particular photometric data. The Monte-Carlo simulations we have used to test the selection bias, completeness and false detection rate of our detection algorithm are described in Section \ref{MC}. As a further test of our method, we compare our cluster detections from the $4 \sqdeg$ of CFHTLS-Deep imaging with those in three other published catalogs,  the XMM-LSS\footnote{XMM-Newton (X-ray Multi Mirror) Large Scale Structure survey} clusters \citep{Pier07}, the catalogs from the photometric redshift based Adaptive Filter method \citep{Mazu07} as well as cluster candidates from the Matched Filter catalog \citep{Olse07}; results from these comparisons are presented in Section \ref{CCDeep}. We present our cluster catalogs from the $161 \sqdeg$ of CFHTLS-W imaging, with each candidate categorized into high, medium and low quality based on the detection likelihood, in Section \ref{CCW3}. We summarize all our results in Section \ref{conc} and outline improvements in progress and future plans. We use Five-Year \emph{WMAP} values \citep{Dunk09, Koma09} of the cosmological parameters throughout.    


\section {Optical detection of clusters} \label{RevCS}

\indent Optically selected cluster catalogs spanning ever increasing redshift intervals continue to be published using innovative detection techniques that have kept pace with the advances in optical observational technology.  In addition, cluster catalogs may also be constructed using detections at other wavelengths, e.g., with X-rays emitted by the hot gas trapped in the deep gravitational potential of clusters (\eg \citet{Paca07}), as well as in microwave wavelengths using the thermal Sunyaev-Zel'dovich effect \citep{LaRo03}, to cite just two; however, we restrict our present discussion to only optical methods. \emph{K2} draws upon the strengths of several earlier methods while addressing their shortcomings, and the descriptions and discussions in this section explain our motivations for devising this new method and to acknowledge existing techniques which we have built into its design. Readers not needing this background may skip directly to Section \ref{ClusSearch}. 

\indent Optical detection methods identify clusters utilizing one or more of the characteristic properties of their member galaxies to effectively isolate them from their field galaxy counterparts. Amongst these, the enhanced density of \emph{bright} galaxies is the most visible marker of the presence of a cluster. Used by both Abell and Zwicky in their pioneering work on visual cluster detection \citep{Abel58, Zwic68}, this density enhancement is used by the automated \emph{Counts in Cells} \citep{Lidm96} and \emph{Vornoi Tessellation} \citep{Kim02,Rame01} methods to effectively isolate candidate clusters in single filter imaging, However, since these methods rely only on position information, they are prone to contamination due to projection effects from foreground and background objects, making them a poor choice for surveys, such as the CFHTLS-W, for which multi-filter data are available. 

\indent This enhanced galaxy density contrast between cluster and field environments is further accentuated by the `Morphology - Density', $(T - \Sigma)$ relation - there is an increasing fraction of early type (E+SO) galaxies in the higher density cores of clusters while the late type galaxies are preferentially found in lower density, field environments \citep{Dres80}. Recent surveys have confirmed this morphological segregation in clusters even at redshift 1 and above \citep{Hels03, Smit05}. Further, these surveys have also found an evolution in the $(T - \Sigma)$ relation with redshift, with the early type galaxy fraction, $f_{(E+S0)} = 0.7$ being lower at $z=1$ compared to the value of 0.9 observed in the local universe. This evolution in the E+S0 fraction is consistent with the `Butcher-Oemler' effect \citep{Butc84} in which high redshift clusters are observed to have an excess of blue galaxies in their cores, indicative of ongoing star formation compared to their low redshift counterparts \citep{Gerk07}. In combination, these two effects manifest themselves as a higher density of (E+S0) galaxies in the cores of clusters compared to the field, with the colors of these galaxies evolving toward the red with advancing cosmic time. 

\indent Based on this model of a concentration of bright (E+S0) galaxies in the cores of clusters, the \emph{Matched Filter} (MF) \citep{Post96} applies a filter which combines the expected (Schechter) luminosity function of cluster members as well as their radial (circularly symmetric) distribution to either a pre-defined grid or to individual galaxies (above a magnitude threshold) in imaging data and detects clusters as enhancements in the resulting density map. In the Matched Filter method, the reliance on an `ideal' cluster model is both a strength and a drawback. By fitting a fiducial cluster model to the data, the method derives useful cluster parameters, such as the richness and redshift, as part of the detection process. The method is ideal for deep, single filter data and can be tuned for different cluster models. At the same time however, the dependency on a cluster model biases the method against clusters which do not fit the assumed profile, e.g. a Bautz-Morgan Type III cluster \citep{Baut70}, without a central galaxy, may be missed by the MF method. This drawback becomes greater at higher redshifts, where the higher merger rates produce significant differences among clusters \citep{Cohn05}. In addition, this method too is prone to projection effects since it fits all galaxies, including projected field galaxies that lie within the aperture, to the assumed cluster model. This contamination may be significantly reduced by including available color information, a technique that is used by the methods described in the following paragraph, and by our detection algorithm as well.  

\indent In addition to the segregation by morphology, the (E+S0) galaxy population in a cluster, especially those within the central core regions, shows a strong correlation between color and absolute magnitude. This \emph{Color-Magnitude Relation} (CMR), \citep{deVa61, Visv77, Bowe92, Stan98, Koda98, Blak03, deLu07}) is observed as a tight ridge, the \emph{Red Sequence} \citep{Glad00}, which extends over several magnitudes with small scatter ($\leq 0.1$mag) in a color-magnitude diagram (CMD); recent results using near infra-red observations indicate that the CMR may have already been in place in proto-clusters as early as $z \sim 3$ \citep{Koda07}. There is an emerging consensus that the observed tight Red Sequence indicates that the stellar populations in cluster ellipticals are homogeneous, formed very early $(z \gg 1)$ and have been passively evolving since then. The observational evidence clearly shows that the CMR is universal, being present even in poor clusters and is also identical in zero point and slope -- within photometric uncertainties -- in any two clusters at the same redshift \citep{Glad00, Koda07}. In addition, brighter, and hence more massive galaxies (assuming a simple mass-to-light ratio), are observed to have redder colors than less massive ones, which therefore leads to a small slope of $\sim0.1$ mag/mag in this color sequence \citep{Koda98}.

\indent Therefore, searching for clusters in color space by using the Red Sequence of the (E+S0) cluster members offers a complementary avenue to reduce the dependence on an ideal cluster model (based only on the magnitude and distribution of the member galaxies) and the consequent selection bias. Such a color-based approach was first implemented by \citet{Glad00} in the \emph{Cluster Red Sequence} (CRS) method, which identifies clusters as \emph{simultaneous} overdensities in a combined (color+position) space. The overdensity in color space due to the clustering of member galaxies along the Red Sequence is isolated using a series of overlapping color slices on a color-magnitude diagram; the slope of each color slice is matched to that of a fiducial Red Sequence in a chosen set of redshift intervals, while the width of each slice is set by the photometric error of the available imaging. This color+position approach efficiently isolates clusters, obtains their (photometric) redshift as part of the detection process and also effectively suppresses foreground and background contamination independent of the redshift of the candidate cluster. With the need for color information however, such methods are restricted to surveys in which the imaging is available in two and preferably more filters. The CRS has been implemented in the ongoing Red Sequence Cluster Survey (RCS), a dedicated cluster survey, which uses a matched pair of $R_c$ and $z'$ filters to isolate the strong $4000 \angs$ break in early type galaxies at redshifts up to 1 \citep{Glad05}.

\indent An alternate color based approach, the \emph{Cut and Enhance} (CE, \citet{Goto02}) designed for SDSS imaging, applies a series of color cuts to isolate subsets of galaxies with similar colors from three filter (Sloan \emph{gri}) photometric catalogs; this selection based on color clustering is augmented with the galaxy number density computed within cells on a grid. The color-position clustering signals are amplified by using a weighting scheme to create density maps on which `Sextractor' \citep{Bert96} is used to detect peaks as candidate clusters. 

\indent In the closely related \emph{MaxBCG} method \citep{Hans05}, also developed to work with SDSS imaging, each galaxy is tested for the likelihood of being a Bright Cluster Galaxy (BCG). The method uses a combination of the number of galaxies that lie within a defined color slice of the galaxy being tested (thus checking for a Red Sequence) and a fit of the galaxy's properties to an \emph{empirical} BCG evolutionary track. The likelihood function is tested at a series of redshift slices and maxima in the likelihood function are used to locate BCGs and the clusters in which they occur. Based on a BCG evolutionary model, this method obtains the richness and redshift of the cluster as part of the fit. It must be pointed out however that recent results \citep{deLu07a,Bild07} indicate that BCGs in certain rich clusters have ongoing star formation with colors up to $\sim1$mag bluer than the red sequence; such clusters would not be identified by the MaxBCG. In addition, like the Matched Filter method, the MaxBCG too would miss clusters which lack a central galaxy, i.e., clusters classified as Bautz-Morgan Type III clusters \citep{Baut70}. Finally, the dependency on an empirical evolutionary track for all BCGs adds a further selection bias.

\indent Knowledge of the redshift of even a subset of the galaxies in the survey area provides a third dimension and a very stringent constraint that may be employed for cluster detection. By incorporating available redshifts, the \emph{Adaptive Matched Filter} (AMF) retains all the powers of the basic Matched Filter algorithm \citep{Post96} while significantly reducing or removing noise from foreground and background contamination; the method also provides better estimates of cluster membership and richness. However, the dependency on a cluster model still leaves it prone to miss clusters which do not match the input model, \citep{Kim02}. The \emph{C4 algorithm} \citep{Mill05}, implemented on 2600$\,\sqdeg$ of SDSS data, incorporates position, five filter photometry and \emph{spectroscopic} redshifts to identify clusters at over 90\% completeness and 95\% purity for clusters with mass $\geq 10^{14} \msun$ and up to redshift 0.12, as shown by extensive Monte-Carlo tests. The main challenge in applying this method for detecting clusters at higher redshift is the dependency on spectroscopic redshifts, which at present is expensive in terms of observing time. 

\indent \emph{Photometric} redshifts (\emph{photoz}), estimated from multi-color imaging, offer a viable alternative to utilize redshift information for cluster detection without incurring the penalty of long exposure times. \citet{Mazu07} have established a benchmark for this approach using the public photometric redshift catalog for CFHTLS-Deep 1 field, estimated from multi-color optical and near infra-red imaging \citep{Ilbe06}. For cluster detection, \citet{Mazu07} use an adaptive kernel to locate density enhancements in $i$-band imaging in overlapping photometric redshift slices of width $\Delta\,z = 0.1$. In a comparison with the XMM-LSS cluster catalog for this field \citep{Pier07}, the method is shown to recover $100\%$ of the X-ray detections. Given the overlap of CFHTLS-D1 with our search area, we compare the \emph{K2} detections with those of both the XMM-LSS \citep{Pier07} and Adaptive Kernel \citep{Mazu07} catalogs in Section \ref{CCDeep}, and therefore provide further pertinent details of both these catalogs in the corresponding subsections.

\indent In an alternative approach using photometric redshifts, \citet{Gava07} combine weak lensing reconstructions in the $i$-band images with the available photoz catalogs for the four CFHTLS-Deep fields to identify clusters as peaks in the convergence maps. Comparison with the overlapping XMM-LSS catalogs indicates that nine of the fourteen X-ray clusters are secure detections with redshifts and velocity dispersions being measured as part of the detection process. \citet{Berg08} use a similar approach but use \emph{wavelets} for reconstructing the weak lensing maps; these maps are then combined with CFHTLS-Deep photometric redshift catalogs for cluster detection. In closing, it must be emphasized that even though photometric redshift estimation may not be as expensive in integration time as spectroscopic redshifts, these methods still require \emph{deep, multi-band} ($\geq 5$) imaging, while our method, \emph{K2}, is designed to rely on just two colors and may thus find wider application.


\section {\emph{K2}, a cluster detector for multi-color CFHTLS-W imaging} \label{ClusSearch}

\indent Our detection method, \emph{K2}, is designed to fully utilize the galaxy positions as well as their available colors in processed CFHTLS-W photometric catalogs. The principal idea is to identify simultaneous density enhancements in the projected 2D positions of the galaxies as well as in two colors independently; this method may be trivially extended to include more colors, if available. Cluster candidates are identified using a well defined metric computed with the position-color enhancement. Our search algorithm draws upon the strengths of four other methods reviewed in Section \ref{RevCS}, with improvements tailored for the CFHTLS-W data: -1- the \emph{C4} \citep{Mill05} though without redshift information, -2- to the \emph{Cut and Enhance} (CE)  \citep{Goto02}, with a galaxy based density enhancement estimate, -3- the \emph{MaxBCG} \citep{Hans05} without relying on a BCG evolution model and -4- to the \emph{Cluster Red Sequence} (CRS) \citep{Glad00} though we stipulate a minimum of two colors to increase the robustness of the detections. Our method \emph{does not} impose a luminosity distribution or a modeled radial distribution of cluster galaxies, thereby minimizing any selection bias in the detections. In addition, we found that by locating the detection filter at each galaxy and not on a uniform grid defined over the survey area (as is done in both CE and in CRS methods), our approach becomes more efficient in rejecting false positives due to noise peaks. This galaxy centered approach instead of a positional grid has been adopted in both the \emph{Adaptive Matched Filter} \citep{Kepn99} and the \emph{MaxBCG} \citep{Hans05} methods specifically for this advantage.

\indent At the time of development of our cluster detection method, only \emph{gri} photometry was available for most of CFHTLS-Wide survey area. To test the feasibility of obtaining photometric redshifts with just these two colors, we compared SDSS spectroscopic redshifts (available for bright, $r \leq 17\,$mag, galaxies in a fraction of the Wide fields) with photometric redshift estimates from \emph{HyperZ} \citep{Bolz00} using only the available \emph{gri} magnitudes and errors. However, with only two colors  the photometric redshift estimates for $\sim30 \%$ of the galaxies failed catastrophically. Hence \emph{K2} has been developed to work with just two color imaging and without reliance on redshift information; however, details of plans to incorporate redshift information in our algorithm are discussed in Section \ref{conc}.  

\indent \emph{K2} is designed to work with processed photometric catalogs from multi-filter, wide field imaging. At present, we are using photometric catalogs for CFHTLS-W imaging which are available as part of the \emph{Terapix T05} data release (August 2008); Table \ref{Tbl-1} lists the numbers and other details of the available catalogs for the four survey fields of CFHTLS-W. The Terapix catalogs have been produced with \emph{Sextractor} \citep{Bert96} on the individual stacked image for each field in each filter. During this processing, the \emph{i}-image is used as the reference frame for source extraction from images in the other filters; the Terapix website\footnotemark[3] provides detailed explanations of all the image processing, astrometric and photometric calibrations during the production of the CFHTLS T05 catalogs. Since our cluster detection algorithm depends on the colors of galaxies, we independently compared the photometry of objects found in common between the CFHTLS-W fields and the SDSS imaging catalog; the differences between the photometry in these two catalogs were consistently smaller than the photometric errors, thus confirming that the Terapix photometric catalogs were suitable for our application. 

\begin{figure} 
\epsscale{.80}
\plotone{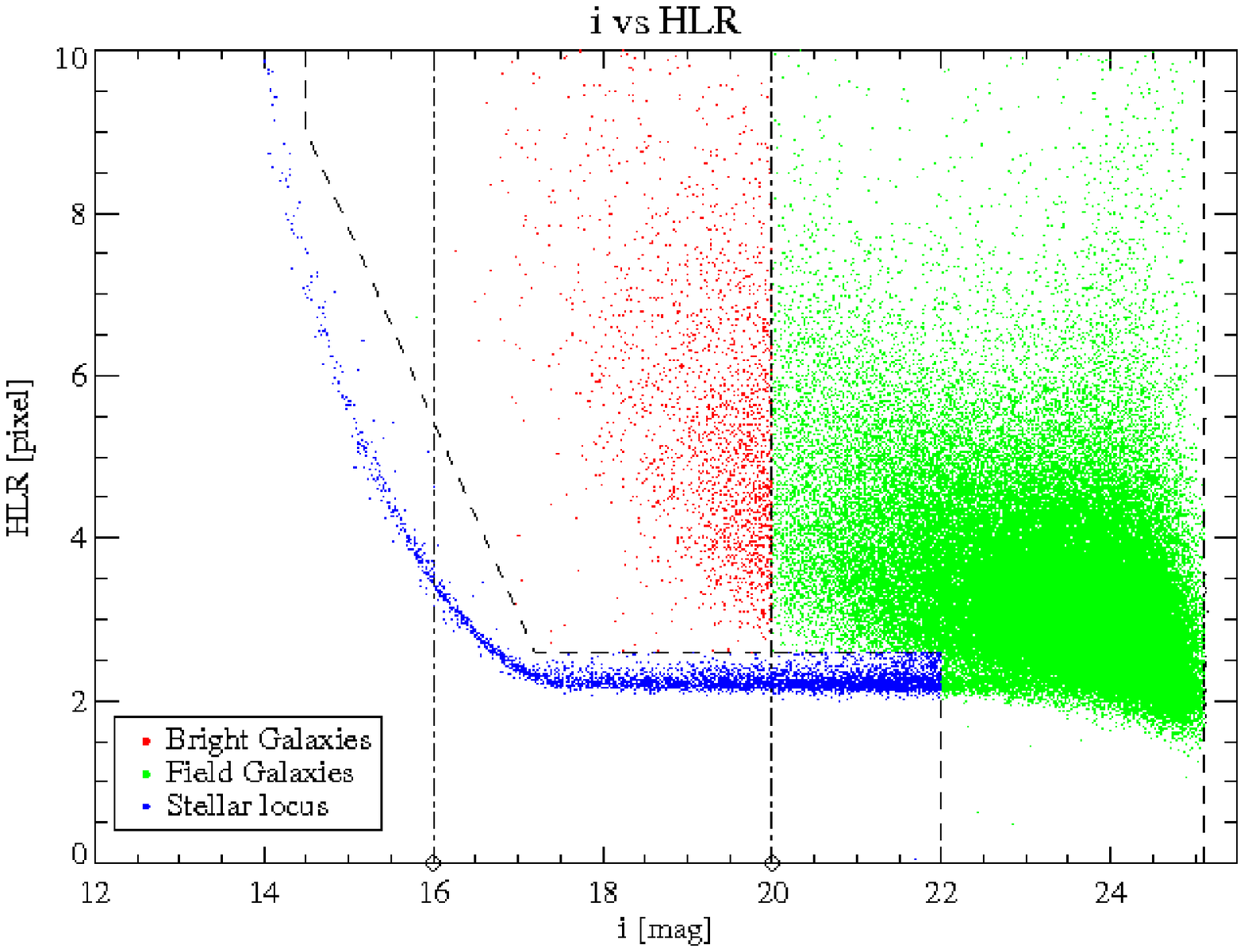}
 \caption{Plot of \emph{i} magnitude versus half light radius (HLR) of all objects in a $1 \sqdeg$ CFHTLS-W field. The stellar locus defined by the HLR is used for star-galaxy segregation. The user defined magnitude and HLR selection cuts (described in \S \ref{ClusSearch}) are marked in dashed lines and the classification identifies stars (in blue), bright galaxies (red dots) and faint field galaxies (in green). Over-plotted are the magnitude cuts (chained lines) used for selecting the bright galaxies (BG), which are tested for cluster membership. \label{Fig1}}
\end{figure}

\indent For each CFHTLS-W field, these Terapix \emph{gri} photometric catalogs form the input to \emph{K2}. For each object in the catalogs, we extract the ID number, sky positions $(\alpha , \delta)$ in J2000 coordinates, x,y positions on the (20k x 20k) image, the \emph{i}-band half-light radius (HLR), and \emph{gri} Sextractor Kron magnitudes with corresponding photometric errors; the Kron photometric apertures in all filters are matched with those of the \emph{i}-band image, which is used for object detection and segmentation. In order to exclude objects with poor photometry in one or more filters, we exclude all objects fainter than a user defined threshold or which have photometric errors greater then a user defined maximum; for CFHTLS-W, these selection limits are $i \leq 24\,$mag and magnitude errors $\leq 0.5$ in \emph{all} three filters, both limits being set to match the depth of the survey photometry. We then use the (\emph{i}  vs HLR) relation, a sample of which is shown in Figure \ref{Fig1}, to isolate galaxies from stellar objects; the blue points, which lie below a HLR threshold demarcate the locus of point-like stellar objects. This HLR threshold, which is dependent on the PSF, is interactively selected for each image frame to account for field-to-field PSF variations; for the CFHTLS-W imaging, the HLR threshold is nominally $\sim0.6\asec$, while the upper limit is set at $4\asec.5$. The magnitude limit for this star -- galaxy discrimination using the HLR method is also set by the user and is nominally $i\,=\,21.5$mag; objects fainter than this limit are taken to be part of the galaxy population. Within the galaxy population, we select only a magnitude limited set of galaxies, referred as \emph{Bright Galaxies} (BG), which we test for cluster membership; we use this magnitude selection to avoid saturated objects and those for which the photometric color errors are greater than the color cuts used for cluster member selection, as explained later in this section. For the CFHTLS-W, the bright end limit for the BG sample is set at $i \geq 16\,$mag, while the faint end is $i \leq 20\,$mag; as with the HLR limits, these magnitude limits too may be varied interactively to take into account variations in the photometry between fields. The pertinent photometric and astrometric details of the BG are then written to a \emph{BG catalog} for the cluster detection stage to follow. Similarly, a \emph{field catalog} is prepared consisting of \emph{all} objects in the observed field, which meet the magnitude and photometric selection criteria mentioned above. Each 1 $\mathrm{deg^2}$ CFHTLS-Wide field typically contains of order 2000 BGs and 150,000 field objects. 

\indent For each BG in the catalog, \emph{K2} computes the (position+color) enhancement (referred as the \emph{weight}) within a predefined aperture centered on the galaxy; \emph{the weight in each color is computed independently and the results are combined only in the last step of the detection process, when candidate cluster members are linked to identify clusters}. For each BG, we select \emph{all} objects (field objects as well as other BGs) within a fixed aperture of angular size of $82\asec$, which corresponds to a radius $0.5\,\mpc$ at redshift $z = 0.5$; this physical radius corresponds to half the virial radius of a typical cluster of virial mass, (M$\sim10^{14}\,\msun$). The chosen aperture size therefore adequately covers the cluster core from which (E+S0) galaxies contribute the majority of the position and color weight\footnote{We also tested the feasibility of basing aperture size on the photometric redshift of the BG; however, due to errors in the photometric redshifts evaluated using \emph{only two} colors, this adaptive aperture has not been implemented in the current version of \emph{K2}. However, we aim to incorporate this approach in a future version tuned for five filter imaging, as discussed in \S \ref{conc}}. Other detection methods, \eg  \citet{Hans05, Kim02}, adopt a similar detection aperture size for this reason. 

\indent For the set of objects within the aperture, we then apply a color cut centered on the color of the BG to locate likely co-cluster members which lie along the Red Sequence. The width of the color cut, ($= \pm 0.15\,$mag), corresponds to the photometric color error of a galaxy at the faint magnitude end ($i = 24\,$mag) of our field sample. The \emph{cluster weight}, as a measure of the BG being a cluster member (our null hypothesis),  is computed as, 
\begin{equation}
\label{eq:wteqn}
W_c = \sum_{i=1}^n \frac{1}{(\delta_d + \epsilon_d)} \ast \frac{1}{(\delta_c^2 + \epsilon_c)}  \end{equation}
where the position and color separations, $\delta$, are measured with respect to the BG; suffixes refer to the separation in distance (d, measured in pixels) and in color (c, in magnitude). The summation is carried out only for the objects which lie within the aperture as well as within the imposed color separation from the BG; these position and color cuts normally yield a few tens of objects per BG. Softening parameters, $\epsilon$ are used to prevent numerical errors due to division by zero; the softening parameter for distance,  $\epsilon_d = 0.02\,$pixel (a tenth of the astrometric error) and for color, $\epsilon_c = 0.005\,$mag (a tenth of the minimum color error for a BG ). The softening parameters thus represent a small fraction of the smallest expected error values in position and color.  

\indent Next, we compute the \emph{field weight}, which is a measure of the BG being just a field galaxy (the alternate hypothesis). For this we use the field catalog with shuffled x, y positions for all the objects; their colors remain unchanged. The BG is placed at 1000 random positions in this shuffled field and at each position the weight, given by Equation \ref{eq:wteqn}, is computed. The median and the inter-quartile distance (IQD) of this set of 1000 weights is taken to be the field weight, $W_f$, and a measure of the scatter, $\sigma_f$. Finally, the \emph{significance}, which is the metric that we use to measure the likelihood of the BG being a cluster member, is calculated by,
\begin{equation}
\label{eq:sigeqn}
S_c = \frac{W_c - W_f}{\sigma_f}
\end{equation}
We repeat this computation for all the galaxies in the BG catalog. 

\indent Detected cluster members are those with significance values greater than a threshold, ($3 \,\sigma$ in the current implementation), in \emph{both} the (\emph{g-r}) and the (\emph{r-i}) colors. This threshold value was determined using the detection efficiency from our Monte-Carlo simulation results, described in \S \ref{MCres}. 

\indent In the preceding steps, each BG is tested individually for cluster membership; subsets of these BGs may be members of the same clusters. We therefore group together individual cluster members (BGs with significance greater than the threshold) using a `Friends-of-friends' algorithm \citep{Huch82} in which galaxies within a certain linking length are combined into a single group. We use both colors and physical separations for linking member galaxies; our chosen link lengths are the widths of the color cuts and the radius of the detection aperture respectively. For clusters with multiple members, we identify the Bright Central Galaxy (BCG) as the member with the brightest $i$ magnitude; for candidates in which only a single BG lies above the magnitude cut, that galaxy is also taken to be the BCG and its properties are catalogued as described in \S \ref{CCW3}. Finally, we separate the detected clusters further into  \emph{Gold, Silver} and \emph{Bronze} categories depending on their maximum detection significances being greater than $5 \,\sigma$ in both colors, greater than $5 \,\sigma$ in just one color or less than $5 \,\sigma$ in both, respectively; note that all these cluster members satisfy the minimum detection significance of $3 \,\sigma$, as mentioned earlier.


\subsection{Salient features of \emph{K2} \label{K2feat}}
Before discussing the operational characteristics of \emph{K2}, we highlight salient features which have been specifically designed to make the method flexible, portable and minimally dependent on a fiducial cluster model. 

\indent \emph{K2} has the built in flexibility to accept photometric catalogs in as many imaging colors as available, the minimum being two; the greater the number of colors used, the higher is the overall detection significance, since only the combined significances (a minimum of $3\sigma$) in all colors flags a detection, as explained above. In addition, the results reported in this paper are from the current implementation on the CFHTLS-W $gri$ photometric catalogs. However, \emph{K2} is designed to accept photometric catalogs from any similar imaging survey, provided the necessary details ($\eg$, HLR, sky positions, magnitudes) are available.   

\indent For ease of portability, our cluster detection routines are a group of IDL\footnote{Integrated Development Language} procedures packaged into a single \emph{IDL project} file. The complete cluster detection sequence and preparation of cluster catalogs for each $1\,\sqdeg$ CFHTLS-W field takes $\sim 15$ minutes on a single processor PC. The only user intervention needed during execution is to verify the star-galaxy discrimination step, similar to the one shown in Figure \ref{Fig1}, where the default HLR and magnitude limits may be overridden to take into account field-to-field variations in the PSF and photometry.

\indent It is to ensure that the dependency on a fiducial cluster model is kept to a minimum that we utilize the two-step detection process, testing only \emph{individual} galaxies in Step 1 to identify likely cluster members using the combined \emph{significances} in two colors. The operational definition of a \emph{cluster member} is an enhanced density of galaxies within the detection aperture with colors matched (within the width of the color cut) to the color of the galaxy being tested. Only in Step 2 are the identified cluster member candidates then linked together, again using matched colors and the detection aperture as linking length, to identify member galaxies which belong to the same cluster. 

\indent Therefore, \emph{K2} is capable of identifying cluster members with enhancements in \emph{any} color and does not depend on the presence of a `Red Sequence', as does the \emph{CRS} method \citep{Glad00}. The method is thus capable of detecting even spiral rich structures such as `Hercules cluster', Abell2215 at z=0.036, \citep{Stru99} and the `Blue Sequence' reported in the Virgo Cluster by \citet{Bose06}. The detection significances of such clusters will consequently be higher only in blue colors, implying that ($u-g$) and ($g-r$) colors are better suited for the detection of such `blue' sequences.

\indent The above discussion may be extended to clusters in which a `Red Sequence' is present but with a slope that is different from the typical values given by \citet{Koda98}; an example is reported by \citet{Adam07} from deep ($B = 25.25\,$mag) observations of one quadrant of the Coma cluster. In \emph{K2}, the width of the color cut is set equal to a pre-factor $\times$ the fiducial slope of the CMR $\times$ the magnitude range used for detection; the pre-factor (default =1.5) is used to compensate for the associated photometric errors. The default color cut values implemented at present take into account typical values of the CMR slope based on the results from \citet{Koda98}. It must be emphasized that the detection method \emph{does not} fit the slope of the Red Sequence, but merely identifies galaxies within the color cut. In order to identify clusters with CMR slopes with are outliers, such as the one reported by \citet{Adam07}, an increased pre-factor and greater width of the color cut may be used. However, increasing the width of the color will consequently increase the contamination from foreground and background galaxies in the majority of clusters with \emph{typical CMR slopes} and thus lower their detection significances. Fine tuning of this parameter must therefore be assessed against the target cluster population and the intended application of the output catalog. 

\indent A related comment applies to the effect of uncertainties in the measured magnitudes and colors of galaxies on the \emph{K2} detection efficiency. Since \emph{K2} computes the \emph{significance} (Eq \ref{eq:sigeqn}) as a relative measure using cluster and field weights (both of which are affected equally by color errors), it is relatively insensitive to the method by which errors are estimated. For a given photometric catalog, the width of the color cut is chosen to match the color error of a galaxy at the faint end of the magnitude range, as mentioned in Section \ref{ClusSearch}; the default color cut of $\pm0.3\,$mag applies to the current implementation on \emph{Terapix} catalogs. 
 
\indent Finally, \emph{K2} places a detection aperture on each BG and computes the weights, given by Eq \ref{eq:wteqn}, using only the \emph{proximity} of galaxies within the aperture and \emph{not} by matching any radial profile. Since identified BGs are then linked in Step 2 using matched colors and the detection aperture as linking length, the overall \emph{effective} aperture is much greater for multi-BG clusters than the single detection aperture. Thus, this Friends-of-friends linking process is capable of identifying clusters whose members are not distributed according to a classic profile, merging clusters with an extended distribution of members and of recovering filamentary structures such as the one observed in the infalling region in Abell 85 at z = 0.055 \citep{Boue08}. If the colors of the member galaxies in such interacting clusters are different (greater than the width of the color cut), each is identified as an individual structure. Since \emph{K2} does not rely on a classic cluster profile, it identifies substructures within a cluster as part of the host cluster if their member galaxies' colors match, else as a separate entity provided the BGs in the substructure reach the detection threshold. We verify these design features with Monte Carlo tests as well as using comparisons with published cluster catalogs in the following sections.     


\section{Monte Carlo tests for completeness and contamination \label{MC}}

\indent In the design of our detection method, we have kept the dependency on a cluster model to a minimum, thus decreasing bias against clusters which do not fit any preferential model. In addition, this low dependency on a cluster model minimizes the need for fine tuning of parameters, and thus we are able to run the detection on all CFHTLS-W fields with the minimum of fine tuning.

\indent Despite these safeguards, the method is still prone to selection bias since it assumes the universal presence of the Red Sequence in \emph{all} clusters, including poor clusters with few members. We test this assumption and thus estimate the selection bias using two approaches: first with Monte Carlo tests using synthetic photometric catalogs as described in this section and then with a comparison with three other published CFHTLS cluster catalogs, the results of which are discussed in Section \ref{CCDeep}. In addition, we use these Monte Carlo simulations to estimate the contamination of our cluster catalog from false positives, \eg  chance superpositions of galaxies with matched colors, as well as completeness for a specified cluster richness and redshift. 

\subsection{Simulation methodology \label{MCmeth}}
\indent For our Monte Carlo tests, we generated synthetic photometric catalogs by placing synthetic clusters in a synthetic field of galaxies and stars, to mimic an observed CFHTLS-W field. With \emph{K2}, the detection efficiency depends sensitively on color matches, hence we have ensured that the synthetic BG and field photometric catalogs are self-consistent in colors by using published and well tested empirical relations or theoretical results in constructing the synthetic catalogs. We also investigated the method using synthetic clusters on a shuffled CFHTLS-W field catalog and found that our recovery rate was unrealistically high (100\% even at $z \sim 1$) because of significant differences between the synthetic colors of cluster members and the observed colors of the field galaxies. The following paragraph provides details of generating the synthetic field and BG catalogs used in our tests.

\indent For the synthetic field catalog, we assume a total field galaxy population of 150,000, the median number of galaxies with $i \leq 24\,$mag found in a $1 \sqdeg$ CFHTLS-W imaging field, see \S \ref{ClusSearch}. We first generate the field galaxy number distribution, $n(z,r')$, as a function of redshift and apparent r-magnitude, using the model source galaxy distribution commonly used in weak lensing studies \citep{Brai96} 
\begin{equation}
\label{rVSzeqn}
n(z,r') = \frac{\beta z^2 e^{-(z/z_0)^\beta}}{\Gamma(3/\beta)z^3_0}
\end{equation}
This normalized, parametric fit to observed galaxy distributions in deep redshift surveys is applicable up to redshift $z \sim 1$ and $r$-magnitude range, $ 20 \leq r \leq 24\,$mag, and is thus well matched to our range of interest. We use the same functional form and values given in \citet{Brai96} for the parameter, $z_0 = z_0(r)$ and the constant, $\beta = 1.5$. 

\indent In this field galaxy population, we make the conservative assumption that there are equal numbers of early type E and S0 and late type Sb galaxies. Observations of field galaxy luminosity functions and their evolution with redshift for different galaxy morphologies have however shown that the field population is dominated by late types and that their luminosity functions evolve toward higher luminosity with redshift, while early and intermediate types show little evolutionary trend \citep{Lin99}. Therefore our assumption of equal numbers assigns a higher number density of (E+S0) galaxies to the field than the observed value; consequently, there is an increase in the field weight (more red galaxies in the field matching the color of the BGs) and therefore a reduction in the detection significance; in other words, the detection significance from our Monte Carlo trials may be taken to be lower limits, making them a more stringent test of our detection method. 

\indent For these three galaxy types, we compute the synthetic colors with the evolutionary galaxy spectral energy distribution (SED) models \citep{Bruz03} assuming Scalo IMF \citep{Scal86}, exponentially decaying star formation rates (with timescales of 1-2 Gyr for the early type galaxies and 5 Gyr for the Sb population) and a uniform formation redshift, $z = 10$. We have verified that varying the parameters of the SED models, such as formation redshift or metallicity, have little effect on the results of our simulations since such changes affect both the field and the cluster galaxy colors equally; for our simulations, it is necessary that the field and cluster galaxy colors be consistent \emph{relative} to each other, their absolute values are of little import. These synthetic SEDs are then convolved with the transmission functions of the CFHT \emph{gri} filters. To the generated colors, we add photometric errors using the prescription in Sextractor \citep{Bert96} with the published CFHTLS magnitude limits for the different filters. The positions of the field galaxies are distributed randomly within the (20k$\times$20k) field. 

\indent Our objective is to test the effectiveness of our method in detecting clusters with different numbers of members and Abell richness \citep{Abel58} and at various redshifts. Therefore, for generating each synthetic cluster, we start with a prescribed number of member galaxies and select their absolute magnitudes from the cluster luminosity function. For our simulations, we use published values of Schechter function parameters, $M^*-5log(h) = -20.67$, faint end slope $\alpha = -1.2$ and normalization, $\phi^* = 0.0146\, h^3\,Mpc^{-3}$ \citep{Blan03}. It must be mentioned that the Schechter parameters describing the cluster luminosity function have been observed to vary with the depth of the available imaging, e.g., a steeper faint end slope observed in deep surveys \citep{Zucc06}, as well as in detailed models \citep{Khoc07}. To account for such variations, during the development of our Monte Carlo simulations we used several sets of Schechter parameters, ($-20.2 \leq M^* \leq -21.5$, $-0.7 \leq \alpha \leq -1.25$, $0.01 \leq \phi^* \leq 0.045$) taken from the literature  \citep{Pope05, Zand06, Zucc06, Blan05, Blan03}. However, our results for contamination and completeness of \emph{K2} did not show any significant variations depending on the chosen values of the Schechter parameters. The results presented here therefore are obtained using only the single set of parameters given above. 

\indent Since the objective of these Monte Carlo tests is to evaluate any inherent bias in our detection method only toward galaxy \emph{colors} and not their absolute magnitudes, we take these field galaxy Schechter parameter values to apply equally well to our synthetic cluster member population. With these generated absolute magnitudes, we then compute the Abell richness of the synthetic cluster at its assigned redshift. We iterate this procedure till the generated cluster richness matches the Abell richness value for which the simulation is carried out.

\indent We assign the morphologies and projected positions of member galaxies using observed results from literature for the morphology-density relation and the luminosity segregation of cluster galaxies. Applying the observed luminosity segregation from the ESO Nearby Cluster Survey (ENACS, \citet{Bivi02}), we designate all cluster members with absolute magnitude $M_r \leq -22\,$mag as elliptical galaxies, which populate only the cluster core. Probability density functions, derived from Figure 4 in \citet{Bivi02}, are used to assign the projected radial distribution of these bright galaxies; their angular distribution is randomly assigned, assuming circular symmetry. In keeping with these observational results provided by \citet{Bivi02}, cluster members fainter than $M_r = -22\,$mag are assumed to follow the Plummer radial distribution with a maximum radius, $r_{max} = 1.2\,\Mpc$ and core radius, $r_c = 0.1\,r_{max}$; their angular distribution is assumed to be circularly symmetric and is therefore randomly assigned. Depending on the radial position of each member, the morphology (E, S0 and Sb) is determined using the observed morphology-density results from \citet{Thom06} (the required probability densities are computed from results in Figure 2 of the publication). The colors of the E, S0 and Sb cluster galaxies are obtained from the \emph{same} SED models for these morphologies used for the field galaxies, thus maintaining color consistency.  At the assumed cluster redshift, member galaxies with apparent magnitudes below the limiting magnitude assumed in our work ($i > 24\,$mag) are not included. Photometric errors for the cluster members are assigned using the same prescription as for the field galaxy population, thus maintaining consistency in their photometry. Finally, we wish to emphasize that though we use the Schechter luminosity function and the Plummer radial distribution function to generate the synthetic cluster, our detection method \emph{does not rely} on these models and uses only cuts in position and in colors for locating candidate cluster members.     

\indent The synthetic cluster is then merged with the field by placing it at a random position in the 20k x 20k field and converting the radial and angular positions of the cluster members into field x,y positions. All the cluster members brighter than the BG magnitude cut (see Section \ref{ClusSearch}) form the synthetic BG catalog, which is used for detection; all the cluster members are treated as part of the field catalog as well. These two catalogs form the input to the cluster detector, thus following the methodology we use for the actual CFHTLS-W fields. The detection routine is repeated for a user defined number of trials (typically 100 times) for each synthetic cluster, with a different cluster location for each run. The ($g-r$) and ($r-i$) detection significances from these trials as well as their median and inter quartile values are used for the analyses of the contamination rates and the completeness of our detections These results from the Monte Carlo simulations are discussed in the following section.   

\subsection{Monte Carlo results \label{MCres}}
\begin{figure} 
\epsscale{.80}
\plotone{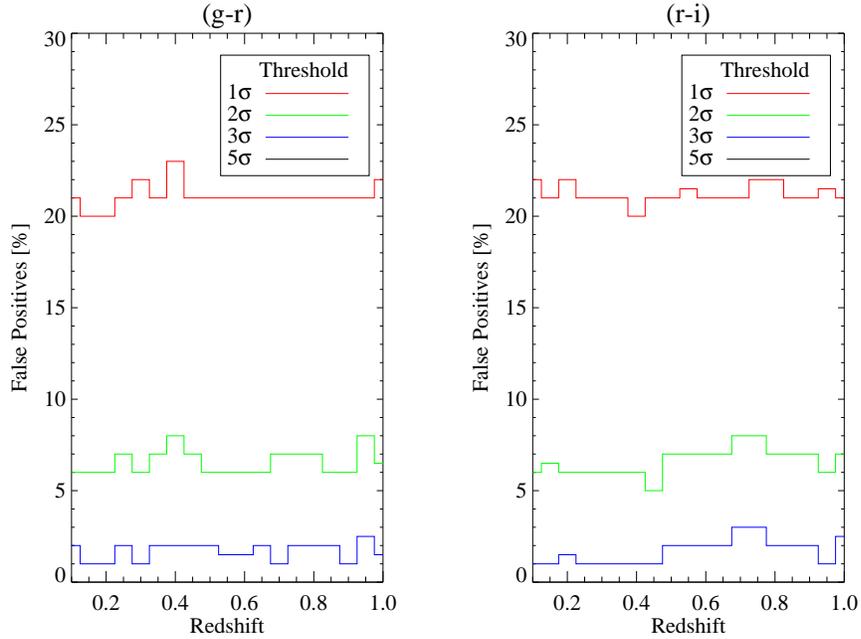}
  \caption[False detection rates as a function of four significance threshold values]{Monte-Carlo estimates of the percentage of false positive detections as a function of redshift  for different detection significance thresholds, as defined in Eq \ref{eq:sigeqn}. The left and right panels refer to the ($g-r$) and ($r-i$) colors used for detection by \emph{K2}. The plotted colors refer to threshold levels of $1 \sigma$ (red),  $2 \sigma$ (green), $3 \sigma$ (blue) and $5 \sigma$ (black, \emph{essentially zero}) \label{Fig2}}
\end{figure}

\indent We first test the rate of false positives for different significance thresholds, (Equation \ref{eq:sigeqn}), in order to check for any underlying bias in our multi-color based cluster detection method, for instance due to a chance superposition of galaxies being classified as a cluster. For this we select 100 random BGs with $i < 20\,$mag and compute the median and scatter in just their \emph{field} weights from 100 random locations in the field, repeating this process in redshift increments of $\Delta z = 0.05$ in the interval $ 0.1 \leq z \leq 1$. The false positive rate is taken to be the percentage number of times a galaxy in a random field position is flagged as a cluster with significance greater than the chosen threshold in a particular color. The histograms, shown separately for the ($g-r$) and the ($r-i$) colors in Figure \ref{Fig2}, are compilations of the median false positive rates for a representative set of galaxies, each bar representing a redshift bin; results for different significance thresholds are represented by the colors indicated. We test the false detection rates for significance thresholds of 1, 2, 3 and 5$\,\sigma$. 

\indent The results show that a $1 \,\sigma$ detection threshold leads to an appreciable contamination due to false positives of $\sim 20 \%$ in both colors; the contamination rate shows no clear correlation with redshift, since the detections are based on galaxy colors and not on magnitudes. The false positive rate decreases to $\sim 5\%$ at a detection threshold of $2 \,\sigma$ and drops to $1 \sim 2 \%$ at $3\,\sigma$; with the detection threshold set at $5 \,\sigma$, the contamination drops essentially to zero. Based on these results, we have chosen a detection threshold of $3 \,\sigma$. In addition, it must be emphasized that, for cluster detection, we impose the more stringent constraint that the significance should exceed the detection threshold in \emph{both} colors; therefore, the $3 \,\sigma$ threshold in both colors reduces the combined contamination rate to $\sim 1 \%$. It must be mentioned that in this False Positives test, we included BGs with various magnitudes up to $i \leq 20\,$mag and did not find any dependency.

\begin{figure} 
\epsscale{.80}
\plotone{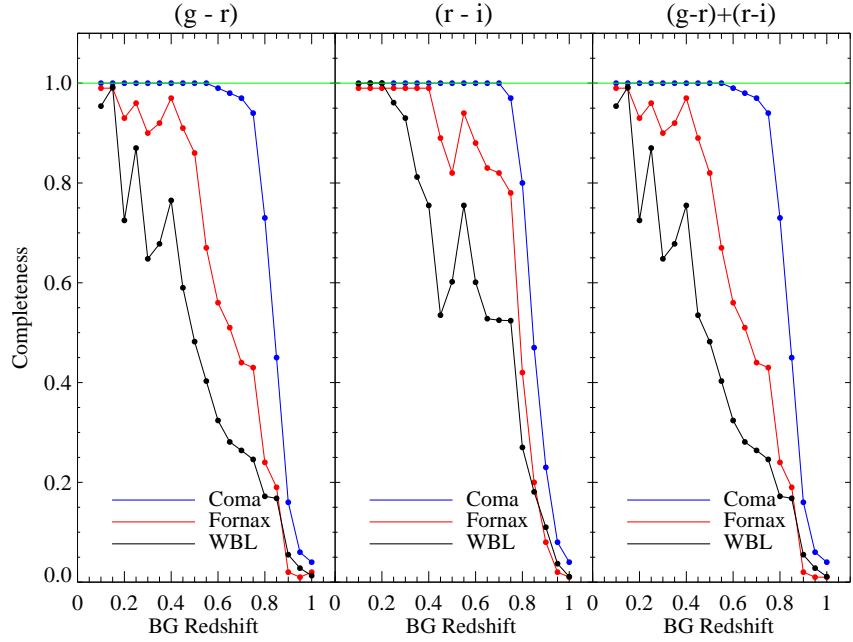}
  \caption{Completeness as a function of redshift for a Coma-like (blue), Fornax-like (red) and a poor WBL cluster (black); the completeness values have been estimated for the (g-r) and (r-i) colors separately, as well as for both colors combined \label{Fig3}}
\end{figure}

\indent As a further test, we compute the completeness of our cluster detections using the recovery rate of synthetic clusters which are generated to resemble three observed clusters of different Abell richness classes. The results presented in Figure \ref{Fig3} pertain to clusters which resemble Coma (Abell richness class $A_c = 2$), Fornax ($A_c = 1$), and a poor WBL cluster ($A_c = 0$) (similar to those in the catalog of poor clusters by White, Blyton and Ledlow, \citet{Whit99}); in the following discussion we refer to these three classes by the names of their template clusters. In the Plummer profile used for the radial distribution of the cluster members (see \S \ref{MCmeth}), the maximum radius for each cluster class is scaled according to the richness, with Coma being assigned $r_{max} = 1.2\,\Mpc$. The results presented here pertain to 100 realizations of each cluster class (ie. 100 Coma-like clusters generated using the method described in \S \ref{MCmeth} and similarly for the other classes). The magnitudes and colors of each cluster as a function of redshift are computed taking into account cosmological dimming and k-corrections, as mentioned earlier; at each redshift, member galaxies fainter than the magnitude limit ($ i > 24\,$mag), are not included in the computation. At each redshift, each cluster is placed at 100 random locations in the field and the detection significance is computed; the Monte Carlo significance for that cluster is taken to be the median of the values obtained from the 100 locations. This computation is repeated for each of the 100 realizations in each cluster class in redshift increments of $\Delta z = 0.05$ in the interval $ 0.1 \leq z \leq 1$. 

\begin{figure}  
\epsscale{.80}
\plotone{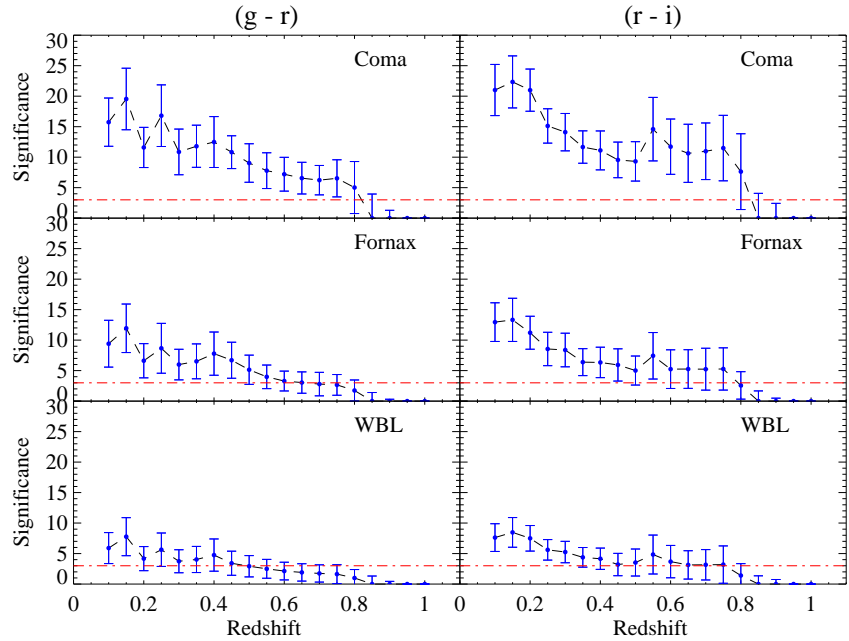}
  \caption{The median and $1 \sigma$ scatter in the detection significance values as a function of redshift computed for the Coma-like, Fornax-like and a poor WBL cluster; the significance values are shown separately for the ($g-r$) and ($r-i$) colors. The red line indicates the $3 \sigma$ detection threshold we have adopted for cluster membership\label{Fig4}}
\end{figure}

\indent Figure \ref{Fig4} plots the median significance values and the $1 \,\sigma$ scatter from the 100 realizations for each cluster class as a function of redshift (each point in that plot therefore represents the median of $10^4$ trials); the left and right panels represent the ($g-r$) and ($r-i$) colors used for detection. Over plotted in each panel is the $3 \,\sigma$ detection threshold we have set using results from the False Positives tests. A comparison of the significances for the three classes at any given redshift shows the expected dominance of the Coma-like clusters with their larger population of (E+S0) galaxies, followed by the Fornax-like and WBL classes. As a function of redshift, the significance of each class trends downward as a greater number of cluster members drop below the magnitude threshold and therefore do not contribute to the computed significance. It is to be noted that the downward trend is not monotonic but shows spikes because the late type galaxies' colors fall within the BG color cut at these small redshift ranges, thus boosting the significance over that at even a slightly lower redshift.  

\indent The completeness of each cluster class in each color at each redshift step is then computed as the fraction of realizations with median significance greater than 3, the threshold we have set for cluster detection based on the false detection estimates. The left and middle panels in Figure \ref{Fig3} show the completeness of our detections in the two colors independently. Since we stipulate that the significance must be greater than $3\,\sigma$ in \emph{both} colors, the plots in rightmost panel show the combined completeness of our method for each of the three cluster classes as functions of redshift. 

\indent The comparison shows that the detections are complete to $ \geq 80 \%$ for Coma in both the ($g-r$) and ($r-i$) colors at all redshifts up to $z \sim 0.8$; the detection efficiency drops rapidly at higher redshifts and asymptotes to a few percent (essentially equal to the false detection rate) by $z \sim 0.9$. This high detection efficiency is driven mainly by the bright early type galaxies in the core; their color clustering increases the detection significance while their bright magnitudes ensure that, even at the higher redshifts, they are above the $i \leq 24$ magnitude limit we have set for our detections. In addition, at all redshifts, the ($r-i$) detection significance is higher, as is expected from the homogeneous red colors of the early type galaxies in the cluster core; the combined detection significance, (rightmost panel), therefore closely mimics the ($g-r$) significance, which is the lower, therefore the more pertinent, of the two significances in determining the completeness. In the other cluster classes, the Fornax clusters are detected at efficiency $ \geq 80 \%$ only to $z \leq 0.6$ while the WBL retain this efficiency only till $z \sim 0.3$. As pointed out earlier, the contribution of late type galaxies to the detection significance leads to the rise and fall in its trend seen in Figure \ref{Fig4}, instead of a monotonic decrease. 

\indent Our principal goal is to identify simultaneous enhancements of galaxy density and color as seen in photometric data and use these as proxies to signal the presence of mass enhancements due to clusters. With the Monte Carlo simulations, we have tested the efficacy of our multi-color based detection scheme and have also determined the optimal detection parameters for this approach.  


\section{Comparison with published CFHTLS cluster catalogs} \label{CCDeep}

As a further test of \emph{K2} as well as to understand any inherent detection bias, we compared our cluster catalogs for  the CFHTLS-Deep fields, which cover four patches each of $1\sqdeg$ within the larger Wide fields, with those of three other published cluster catalogs for the same survey regions, (1) the Matched Filter (MF) cluster catalogs \citep{Olse07} for the four CFHTLS-Deep fields, D1 - D4, (2) the XMM-LSS cluster catalog for the $1 \sqdeg $ CFHTLS-D1 field \citep{Pier07}, and (3) the high likelihood cluster candidates in the Adaptive Kernel (AK) catalogs for CFHTLS-D1 \citep{Mazu07} based on the photometric redshift catalogs for the four Deep fields \citep{Ilbe06}. We discuss the comparison with each catalog in the following subsections; it should be mentioned that both \citet{Olse07} and \citet{Mazu07} have carried out similar comparisons of their optically selected catalogs with the X-ray selected XMM-LSS detections, thus providing a baseline against which we may compare our \emph{K2} candidates. This also helps us quantify the selection bias and completeness of these optical detection methods.

\subsection{Comparison with MF catalogs} \label{MFcatcomp}
\indent \citet{Olse07} have produced their cluster catalogs by implementing the MF method, as originally proposed by \citet{Post96}, to the $4\sqdeg$ covered by the four patches of CFHTLS-Deep fields. The filter uses the Schechter function \citep{Sche76} and a radial profile characterized by a core and cutoff radii \citep{Post96} respectively for the luminosity and the radial distribution of the cluster members in the fiducial cluster model; the values of these model parameters are listed in \citet{Olse07}. They also describe the calibrations of the method using simulated data and provide details of their cluster catalogs for the four CFHTLS-Deep fields. It must be mentioned that \citet{Olse07} use visual inspection of RGB color images to classify the MF cluster candidates as follows: \emph{Class A} candidates show a clear overdensity of galaxies with homogeneous colors; \emph{Class B} show an overdensity but do not exhibit clustering in colors and, \emph{Class C} visually indicate little enhancement either in position or color of their likely member galaxies. \citet{Olse07} do not incorporate any of the available color information in the CFHTLS-D imaging in their detection algorithm, but only base the candidate cluster classification on their RGB colors.

\indent We summarize the MF detection results in Table \ref{Tbl-2}, classified according to their three categories. The numbers listed under these three classes for the MF clusters are taken from their public catalogs\footnote{http://webviz.u-strasbg.fr/viz-bin/VizieR-2, J/A+A/461/81}. Details of the numbers of clusters detected in each of the CFHTLS-Deep fields by our method, \emph{K2}, are also given in Table \ref{Tbl-2} for comparison. The \emph{K2} detections are separated into \emph{Gold, Silver} and \emph{Bronze} categories (as defined in Section \ref{ClusSearch}).  

\indent  The results in Table \ref{Tbl-2} show that there are noticeable field to field variations both in the numbers of clusters detected by each method, as well as in the make-up of the categories of these detections, though the patterns of variations are different. Other than cosmic variance due the presence of large scale structure along the lines of sight of these $1\sqdeg$  patches, these differences may also arise due to variations in the depths of the observations in the three filters in these fields.

\indent Carrying this comparison a step further, Table \ref{Tbl-3} lists the breakdown of each detection class in the MF cluster catalog, which are also detected by our cluster method in this blind trial; the results pertain to the Deep 1 field, with similar breakdown values obtained for the other Deep fields. To generate this comparison, the positions of the candidate clusters in our cluster catalog are matched with those of the MF clusters, with a match being assigned if the positions fall within a distance equal to our detection aperture. It must be noted that during this match up, it was found that $\sim30\%$ of our clusters contain more than one MF cluster; this was specially true of our high significance \emph{Gold} category. The reason is because we link together detected BGs within one detection aperture radius of each other into a single cluster candidate; the MF method on the other hand, treats each detection location as an independent cluster. 

\indent The comparison shows that our detection method matches all the MF Class A detections that lie at redshifts z $\leq 0.8$; the 7 non-detections lie at higher redshifts, according to the photometric redshift value assigned by the MF cluster filter. Similarly, the $z \geq 0.8$ MF candidates in Class B and C are not detected by our method along with $\sim25\%$ of cluster detections at redshift below 0.8; nearly all these low redshift non-detections lie at z $\leq 0.2$. One possible cause may be that our detection aperture size is not including all the galaxies in the core of these low redshift clusters to provide the $3\,\sigma$ minimum detection significance. However, this effect is not observed in the Monte Carlo tests, where we observed the converse in that the detection significance was insensitive to aperture size above a certain minimum radius; this is the justification for the fixed detection aperture we use. An adaptive aperture based on photometric redshift, planned for a future version of cluster detector, may address this cause, if present. The other possible reason for the mismatch between MF and our detections at z $\leq 0.2$ may be that the MF method, relying solely on galaxy density enhancement in a single filter, is more prone to contamination at lower redshifts, where the number of galaxies above the magnitude threshold is higher, leading to enhanced density. 

\indent Visual support of the above arguments is illustrated in Figure \ref{Fig5}, in which the three RGB images in the left column show cluster candidates detected by \emph{both} the MF and \emph{K2} methods (top), by the MF only (middle) and by \emph{K2} only (bottom panel). The top panel clearly shows that a `typical' cluster to the eye is detected by both methods consistently. On the other hand, a cluster with multiple central galaxies or without a radial symmetry (as seen in the lower panel) is missed by the MF perhaps due to the method's reliance on a fiducial cluster model; our detection method, which relies solely on position and color matching of the cluster members, is able to identify such cluster candidates. Our method, however, does not detect high redshift candidates (middle panel), mainly because of the faint magnitudes and associated color errors, which are greater than the color cuts used for selecting member galaxies. These high redshift non-detections match the completeness trends predicted by our Monte Carlo results in Section \ref{MCres}, which show a steep drop in detection efficiency even for a rich Coma-like cluster at $z\geq 0.8$. In these high redshift clusters, the brighter \emph{z}-magnitude of the cluster members may provide better S/N values, and therefore, the present redshift detection boundary of \emph{K2} may be extended to $z \geq 1.$ with the future inclusion of the (\emph{i-z}) color for detection. 

\indent Finally, it must be emphasized that the MF uses only the \emph{i}-filter photometric catalog during the detection process and thus has no color information, even though data for the Deep fields are available in all five filters; the only use of color in the MF method is to \emph{visually} classify their detections into the three classes, as mentioned above. Differences in detection efficiency between the cluster detection methods have been shown to be significant by \citet{Kim02} in their comparison of the Matched Filter method with Vornoi Tessellation, the principal reason being the assumption of a uniform background by the MF scheme. In our scheme, the inclusion of color makes it less susceptible to contamination by the background and this may be the reason for the differences in detection numbers as well as in the assignment of the detection classes.   

\subsection{Comparison with XMM-LSS cluster catalogs} \label{XMMcatcomp}
\begin{figure}  
\epsscale{.80}
\plotone{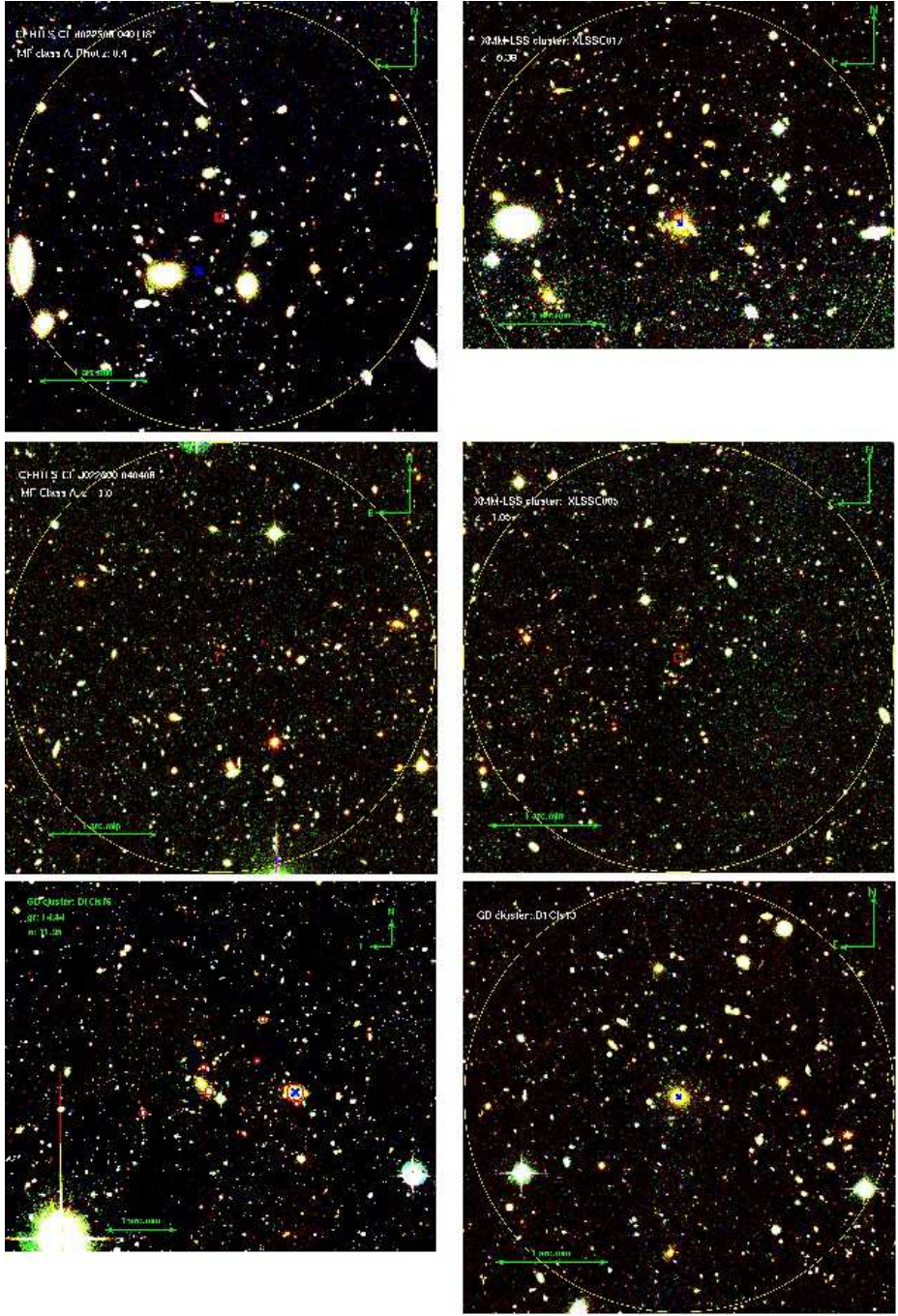}
\caption{Mosaic of RGB color images of representative Matched Filter (\emph{left panels}) and XMM-LSS clusters (\emph{right panels}) matched against cluster candidates in our catalog for CFHTLS Deep-1 field. See \S \ref{CCDeep} for a description of each panel.\label{Fig5}}
\end{figure}

\indent The availability of the \emph{spectroscopically confirmed X-ray clusters} from the XMM-LSS survey region which overlaps the CFHTLS-Deep 1 field \citep{Pier07}, provides a more stringent test for our cluster detection method; since \citet{Olse07} have carried out a similar comparison with their Matched Filter catalogs, we also gain some insight into the detection bias of these two optical methods from these comparisons. We match the cluster detections in our catalogs with the published XMM-LSS cluster positions using the same method as for the MF comparison; a detected cluster BG in our catalog (with $i  \leq 20\,$mag, detection significances in both colors $\geq 3\,\sigma$), whose separation from the published central position of X-ray emission (= position of the X-ray cluster) is less than the aperture radius used by \emph{K2} is taken to be a matched detection. We take these XMM-LSS positions from \citet{Olse07} (Table 5) in order to maintain consistency in our comparison with the MF method. In preparing the cluster catalog for CFHTLS-D1, we use the same \emph{K2} detection parameters as those used normally for all Wide fields and each cluster, if detected, is classified as Gold, Silver and Bronze based on their combined detection significances. The detection results provided in Table \ref{Tbl-4} list whether each XMM cluster is detected by our method (K2) and by the Matched Filter (MF) scheme (as given in \citet{Olse07}) as well as the detection significances in the (\emph{g-r}) and (\emph{r-i}) colors and the corresponding cluster classification. As a visual check of this comparison, the three RGB images in the right column of Figure \ref{Fig5} show, (\emph{top panel}) a representative XMM-LSS cluster also detected by our method, (\emph{middle}) an X-ray cluster not detected by \emph{K2} and (\emph{bottom panel}) one of our \emph{Gold} clusters in Deep-1, which does not have any associated X-ray emission and hence is not part of the XMM-LSS catalog.     

\indent The comparison shows that our method detects 12 of the 17 XMM clusters. It is significant to note that our method detects all the X-ray clusters up to $z=0.8$; all 5 clusters which are not detected are faint, high redshift clusters at redshifts higher than 0.8. As seen in the middle panel of Figure \ref{Fig5}, the faint members of these high redshift clusters are indistinguishable against the field galaxies in the foreground; the \emph{i}-magnitude cut used by \emph{K2} fails even to locate a BG for XLSSC029 at $z=1.05$. On the other hand, the high significance \emph{Gold} cluster in the bottom panel and which visually resembles a `typical' cluster with a BCG and member galaxies with homogeneous colors, is not part of the XMM-LSS catalog, perhaps due to a lack of associated X-ray emission. These results are fully consistent with our results from the Monte-Carlo simulations in which the detection significances in both colors and therefore the completeness are $\geq 80\%$ even for Fornax-like clusters up to redshift 0.6 and then drop steeply off beyond $z = 0.8$ even for a Coma-like cluster with Abell richness = 2. However, as emphasized under Section \ref{MCres}, our \emph{principal} focus at present, using only the available (\emph{g-r}) and (\emph{r-i}) colors, is to identify clusters up to redshift $\sim 0.6$; we aim to extend this to higher redshifts once we incorporate the \emph{z}-filter catalogs into \emph{K2}, as discussed in \S \ref{conc}. 

\indent In comparison, the MF success rate for the XMM clusters is 10 out of 17. It is interesting to note that the MF method fails to detect three clusters at $z\,=\,0.29,\,0.34,\, \mathrm{and} \, 0.46$, all of which are detected with high significance by \emph{K2}, while it successfully detects two higher redshift clusters at $z\,=\,0.92,\,\mathrm{and} \,  1.05$, though with redshift errors of $\sim 0.1$; at the same time the MF fails on a $z\,=\,0.92$ cluster which is detected as a 'Gold' detection by our color based scheme.

\subsection{Comparison with Adaptive Kernel (AK) cluster catalogs} \label{AKcatcomp}
\indent \citet{Mazu07} have used the publicly available photometric redshift catalogs for the four CFHTLS Deep fields \citep{Ilbe06} to develop their cluster detection method based on the Adaptive Kernel technique \citep{Silv86, Dres88} and benchmark their detections for the  $1\sqdeg$ CFHTLS-D1 field against those in the XMM-LSS cluster catalogs for the same area. The \citet{Ilbe06} catalogs provide well characterized photometric redshifts for all galaxies up to $i = 24\,$mag in the four CFHTLS-Deep fields, obtained using a combination of CFHTLS $ugriz$ imaging, along with $BVRI$ photometry from the VIMOS VLT Deep Survey \citep{LeFe05} and infra-red $J$ and $K$-band imaging \citep{Iovi05}. The  error in the photometric redshifts, estimated from a comparison with VVDS spectroscopic redshifts for objects in the regions of overlap with the CFHTLS, is $\pm 0.03$ to 0.07 (depending on the galaxy magnitude) up to z = 1.5 . 

\indent In using the redshift catalog for CFHTLS-D1, for cluster detection \citet{Mazu07} split the galaxy sample into redshift bins, $\Delta z = 0.1$ and apply the Adaptive Kernel to the resulting redshift slices to build density maps. \emph{Sextractor} is then used to detect enhancements in galaxy density in these maps, as well as to obtain the geometric properties (ellipticity, orientation, etc.) of the regions of enhancement. In the 22 redshift slices used, a total of 40 peaks are detected, which include all 8 Class 1 (C1) candidates from the XMM-LSS catalogs \citep{Pier06}. Of these however, only 16 main structures appear in two or more redshift slices (including 4 XMM-LSS clusters) and are therefore identified as high likelihood cluster detections. Of these cluster candidates, 15 also have spectroscopic redshifts by matching member galaxies with overlapping VVDS survey areas.

\indent Having already carried out an independent comparison of \emph{K2} detections with the XMM-LSS X-ray clusters, as described in \S \ref{XMMcatcomp}, we now compare our results with these 16 high likelihood Adaptive Kernel detections; this provides an additional benchmark for \emph{K2} against a different optical detection method based on photometric redshifts. It must be emphasized that the \emph{AK cluster catalog is built using high quality photometric redshifts, utilizing 8 colors plus sky positions for each object}, against just the 2 colors and positions on which \emph{K2} relies. 

\indent The principal details of the AK candidates are listed in Table \ref{Tbl-4a}, reproduced from Table 6 in \citet{Mazu07}; it must be pointed out that each Adaptive Kernel position given in the table refers to the position of the corresponding peak in the density map (determined by \emph{Sextractor}) and not to an identified BCG of the candidate cluster. Therefore, in order to compare detections with \emph{K2}, which relies on galaxy positions, we compute the significance of each BG (refer to Eq \ref{eq:sigeqn} and Section \ref{ClusSearch} for details) within a co-moving aperture of $1\mpc$ at the redshift of the cluster; this aperture size is used by \citet{Mazu07} in the comparison of their catalogs against the XMM-LSS, and we adopt the  same aperture size for consistency, Cluster member candidates are then identified and linked by \emph{K2} using the methods explained in \S \ref{ClusSearch}. The last five columns in Table \ref{Tbl-4a} list whether \emph{K2} detects the candidate, and if so, the significance values in $gr$ and $ri$ colors, the class of the candidate cluster and the positional offset (in arc sec) between the BCG identified by \emph{K2} and the position given in the $AK$ catalog.

\indent The comparison shows that \emph{K2} detects 15 of the 16 \emph{AK} candidates up to the redshift, z $\sim 0.85$. This significant degree of match provides further support of the detection efficacy of \emph{K2} even given the limited color information. At the same time however, this correspondence between the two detection methods is also to be expected given their reliance on the colors of the galaxies, for detection in the case of \emph{K2} and for estimation of redshifts in the case of \emph{AK}. In addition to the \emph{K2} detections identified closest to the positions given in the \emph{AK} catalogs, four apertures contain two \emph{K2} candidates each, while one aperture contains three \emph{K2} detections, These additional candidates have presumably not been segregated by \emph{AK} as individual peaks due to the background threshold set in \emph{Sextractor}  or have not been linked by \emph{K2} due to scatter in their colors. The nature of these multiple detections may only be confirmed by spectroscopic follow up. The only candidate not detected by \emph{K2} is $ID\#\,29$ detected by AK in 2 redshift bins centered at $z\,=\,0.75$ and $0.85$; it must however be pointed out that \emph{the redshift of this candidate is unconfirmed since there is no overlap with the VVDS spectroscopic catalog}. This uncertainty in the redshift of the candidate (likely a high redshift cluster) coupled with the rapid decline in the completeness of \emph{K2} with redshift (seen in Figure \ref{Fig3}), may be the reason for the non-detection by \emph{K2}. These candidates may also be genuine sub-structures identified as a single cluster by the redshift based AK method, but segregated as individual structures due to differences in galaxy colors by \emph{K2}. Finally, the AK candidate, $ID\#\,12$ listed in Table \ref{Tbl-4a}, which has no matching X-ray detection in the XMM-LSS, is also detected by \emph{K2} as a \emph{Silver} class candidate, as further support of the consistency of detections from these two different color based schemes.  

\indent To summarize, the Monte Carlo results show that our cluster detection method is complete ($\geq 80\%$) up to a redshift of 0.6 for clusters of richness matching or higher than Fornax; for poorer clusters, the completeness remains at $\geq 80\%$ until z=0.3 and declines to $\sim60\%$ by z=0.6. These simulation results are supported by the comparison of our catalogs against clusters detected by the optical MF and AK methods as well as the XMM sample of X-ray selected clusters in the CFHTLS-D1. As discussed in \S \ref{conc}, we expect to increase these completeness values even at higher redshifts with the inclusion of the ($i-z$) color in the detection process.


\section{Cluster catalogs for CFHTLS-W fields \label{CCW3}}
We have constructed \emph{K2} cluster catalogs for all $161 \sqdeg$ of CFHTLS-W fields for which \emph{Terapix T05} \emph{g, r} and \emph{i} photometric catalogs are available. Each $1 \sqdeg$ CFHTLS-W field has a corresponding $K2$ catalog, named using the central RA and Dec of the field.  In each catalog, we list the properties of the BCGs of the detected clusters and groups in that field; Section \ref{ClusSearch} explains how we identify the BCG of each object based upon the number of bright cluster members and their $i$-magnitudes. The properties listed in the $K2$ catalogs are the following: a unique ID number for each detection, J2000 sky positions ($\alpha$ and $\delta$), a photometric redshift estimate (see Appendix for details), $i$ magnitude and (\emph{g-r}) and (\emph{r-i}) colors of the BCG, the (\emph{g-r}) and (\emph{r-i}) detection significances, the \emph{K2} detection class, the number of bright cluster members, the Abell richness and the number of member galaxies, $n_{32}$ as well as the limiting magnitude, $m_{32}$, used in computing the Abell richness. Of these, the positional and photometric details are taken directly from the $Terapix$ catalogs, while all the other properties are estimated as part of the detection process. Table \ref{Tbl-5} shows the format of a sample \emph{K2} catalog. $K2$ also lists the boundaries of each cluster candidate (in pixel coordinates on the $20k\times20k$ Megaprime image), which are used internally to generate color RGB images; we carry out the visual search for lenses on these RGB images

\indent In addition to these primary catalogs for all detections, $K2$ also generates an additional catalog for any cluster or group in which multiple BGs have been detected. These complementary catalogs list the astrometric and photometric details of the individual member galaxies as well as their estimated photometric redshifts. Table \ref{Tbl-6} shows a typical catalog of cluster member properties, of which the BCG properties form the first row.

\begin{figure}  
\epsscale{.80}
\plotone{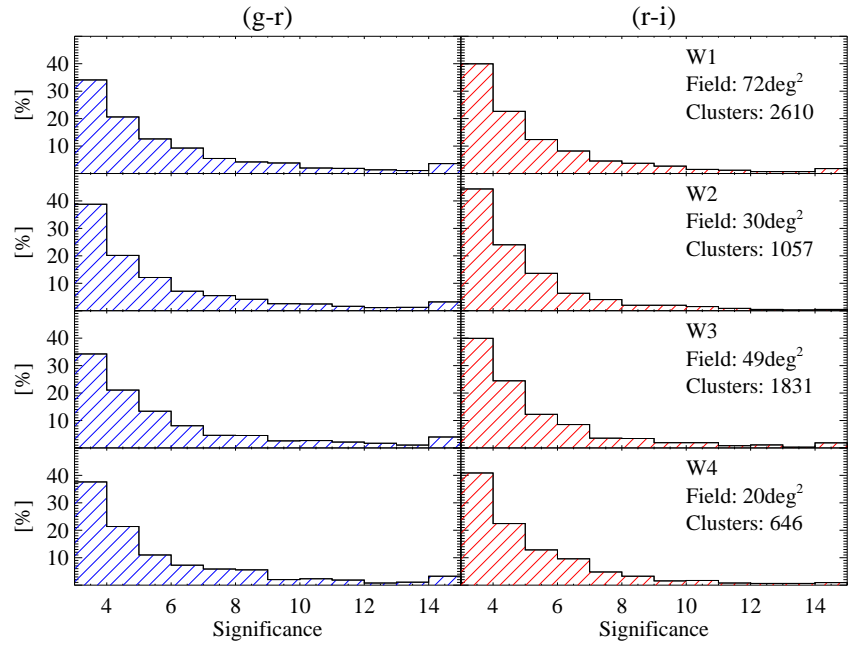}
\caption{Histograms showing the distribution of the $K2$ detection significances of the cluster candidates in the ($g-r$) and ($r-i$) colors, shown independently, in the 161 ${\mathrm deg^2}$ of CFHTLS-W imaging for which $g, r$ and $i$ imaging are available \label{Fig6}}
\end{figure}

\begin{figure}  
\epsscale{.80}
\plotone{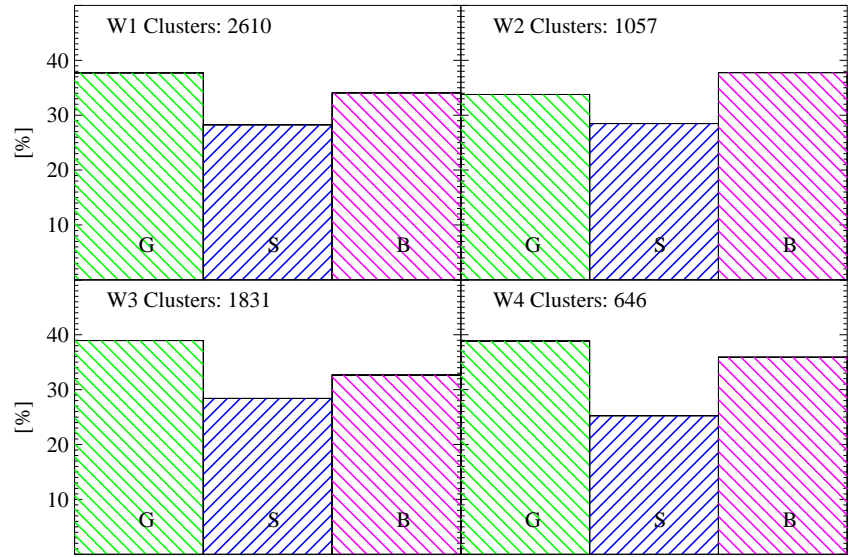}
\caption{Histograms showing the relative percentages of \emph{Gold, Silver} and \emph{Bronze} cluster candidates in the four CFHTLS-W fields. The total number of candidates in each field is annotated for reference. \label{Fig7}}
\end{figure}

\indent Figures \ref{Fig6} - \ref{Fig8} show comparisons of the numbers and characteristics of the detections in the four CFHTLS-W fields. Of these, Figure \ref{Fig6} shows the distribution of the $K2$ detection significances independently in the (\emph{g-r}) and (\emph{r-i}) colors. The histograms run from the minimum $3\,\sigma$ used as the threshold for detection to a maximum of $15\,\sigma$; this maximum value is chosen only for the purpose of the plot, and all objects with significances $\geq 15\,\sigma$ have been included in this bin. It must also be emphasized that an object is flagged as a detection only if it has the minimum $3\,\sigma$ significance in \emph{both} colors; therefore, the (\emph{g-r}) and (\emph{r-i}) histograms for each CFHTLS-W region (corresponding left and right panels in Figure  \ref{Fig6}) contain the same objects, though their significance distributions may differ. The annotated numbers on the right panels indicate the total number of detections in each region. Shown in Figure \ref{Fig7} are comparative histograms of the number of $Gold$, $Silver$ and $Bronze$ detections in each CFHTLS-W survey region; the histogram bins are in percentages of the total number of detections in that Wide field. Figure \ref{Fig8} shows comparative histograms of the Abell richness values of the detections. The bins are labeled according to the generic names we have adopted for the three categories, namely, group scale $WBL$ (W), $Fornax$ (F) and the rich $Coma$ (C) class; the histograms are shown as percentages of the total number of detections in the corresponding survey region. Finally, Table \ref{Tbl-7}, provides a compilation of the quantitative and statistical comparisons of the detections and their classifications in these four survey regions. In the following paragraphs, we discuss these comparative characteristics of the candidates in the four survey regions.

\indent The histograms of detection significances (Figure \ref{Fig6}) show the expected trend of a larger number of candidates \emph{per significance bin} at the lower end of the scale. However, the histograms also illustrate that the \emph{total numbers} of $5\,\sigma$ and greater detections are comparable to the numbers of $3 - 5\,\sigma$ detections. This inference is reaffirmed by the comparison of detection classes in Figure \ref{Fig7} which clearly shows that there are equal numbers of high likelihood $Gold$ objects ($\sim$35\% of the detections) as there are $Bronze$ detections; as explained in \S \ref{ClusSearch}, the classification scheme we use for these detections combines the significances in the (\emph{g-r}) and (\emph{r-i}) colors. 

\indent We next compare the number of detections per square degree in each CFHTLS-W survey region. The median and inter-quartile distances, listed in Row 3 of Table \ref{Tbl-7}, show that there are $\sim35$ detections/$\sqdeg$ (of all three classes) in the CFHTLS-W fields, and that the fields are statistically equivalent. The marginally lower value (30 per $deg^2$) in W4 may be due to cosmic variance since the completeness limits in this region, given in Table \ref{Tbl-1}, match those of the others even though this region was added to the survey two years after its commencement., The small excess number of detections ($\sim 2$ per $\sqdeg$) in the W3 region may also be due to the same reason, with a mass sheet lying along that line of sight. We note that 4 of the 9 lens candidates discovered so far in our search of gravitational lenses in the CFHTLS-W also lie in the W3 region, thus indicating an increased galaxy overdensity in that region. 

\indent The median percentages in the three classes, shown in Figure \ref{Fig7}, indicate that there are nearly equal numbers in the $Gold$ and $Bronze$ categories ($\sim$35\% each), and are greater than the $\sim30$\% in the $Silver$ class; this trend applies to all four Wide fields. This indicates that an object with a detection significance $\geq 5\,\sigma$ in one color is likely also detected with a similar significance in the other color, and vice versa for the lower significance candidates.  

\indent Finally, the distribution of the Abell richness classes of the candidates, Figure \ref{Fig8}, shows that the candidates are predominantly WBL-like objects. With the Abell richness as a proxy for mass, the statistics indicate that galaxy groups and poor clusters form the majority ($\geq 95$\%) of the total detections, or $\geq 30$ per $\sqdeg$. This dominance is seen in all four Wide fields.  Fornax and Coma-like rich clusters make up the remaining 5\%, with a Fornax-class cluster found in every $1-2 \sqdeg$ while a Coma-like cluster occurs every $2-4 \sqdeg$, with W1 showing the highest density (0.49 per $\sqdeg$). The total number of rich clusters in the $K2$ catalogs is 239, which provides us with a rich sample on which to focus our lens search. 


\section{Summary and future direction} \label{conc}
We have developed a new cluster detection method, \emph{K2}, specifically optimized for optical wide field, multi-color imaging. Based on the well tested \emph{Red Sequence} approach, the algorithm uses the observed overdensity in optical colors and positions of early type galaxy population in the cores of galaxy groups and clusters to isolate them from the field galaxy population. The design draws upon select advantages of earlier cluster detection methods with improvements tuned for the available imaging. By testing each galaxy for cluster membership, instead of running the detector on a uniform grid, false detections due to noise peaks or other imaging defects are minimized. Only galaxies in a magnitude limited sample are tested (the limit being set by the depth of photometry available) to improve execution speed and reduce false positives due to large photometric errors. 

\indent $K2$ uses of a well defined $metric$ to compute the \emph{combined} position and color clustering; the metric, and the corresponding detection $significance$ are calculated for each available color independently. Merging these independent lists in the final step of the process, and only selecting detections which lie above a well defined significance threshold, improves the likelihood of the candidates being bona fide clusters. We define a consistent statistical estimate for the $significance$, and thus compute a quantitative measure of the likelihood of a galaxy being a cluster member.  

\indent In order to provide the flexibility to readily incorporate additional colors, $K2$ runs on the photometric catalog for each color independently. Using additional colors, when available, improves the detection significance and the robustness of the detections at higher redshifts. The candidates are classified under $Gold$, $Silver$ and $Bronze$ categories to reflect their likelihood of being genuine clusters. The Abell richness and the photometric redshifts of the cluster candidates are computed as part of the detection process.

\indent Using Monte Carlo simulations, we have determined the optimal detection parameters for $K2$ as well as the completeness and contamination of the resulting cluster catalogs. Based on these tests, we determined that a fixed aperture for computing color and position enhancements around each galaxy is more consistent than an adaptive one based on photometric redshift (this is because our tests indicated that when \emph{only two colors} are used to estimate photometric redshifts of individual galaxies the failure rate is $\sim30\%$). The magnitude limit for the BG selection ($i \leq 20$mag), as well as the number of trials (=1000) used to compute the $field\;weight$ were also determined using these simulations. The Monte Carlo tests have shown that the contamination rate from false detections is less than $1\%$ for the adopted minimum threshold of $3\,\sigma$ for the combined detection significances in both the ($g-r$) and ($r-i$) colors. The Monte-Carlo results have also shown that the completeness of our \emph{K2} catalogs is $\geq 80\%$ for Fornax-like and richer clusters up to a redshift of 0.6, which is the redshift range of interest for searching for cluster lenses, our principal objective; for poor clusters of the WBL-class, the detection method is only complete at this level for redshifts up to 0.3 and drops to $60\%$ by z=0.6. However, the lensing cross section of these low mass systems shows a corresponding steep decline with redshift and the likelihood of detecting lensed images in poor clusters at high redshift is low; therefore the lower completeness does not pose any disadvantage for our application. 

\indent As further verification, the $K2$ catalogs for the CFHTLS-Deep fields have been compared against three published cluster catalogs, namely, the Matched Filter catalog by \citet{Olse07}, the spectroscopically confirmed clusters in the XMM-LSS catalog \citep{Pier07}, and cluster detections by the Adaptive Kernel method for the CFHTLS-D1 field \citep{Mazu07} which are based on photometric redshift catalogs for the CFHTLS-Deep fields by \citet{Ilbe06}. In these comparisons, our candidates match all the confirmed X-ray clusters in the XMM-LSS  up to a redshift of 0.8 as well as 15 of the 16 major structures detected by the Adaptive Kernel method in the same redshift range. At higher redshifts, the decrease in the number of matched detections is because the completeness of $K2$ drops off steeply above $z\sim 0.8$ due to two reasons: (1) the number of member galaxies which lie above the magnitude limit are too few to provide a detection significance above the established threshold value, and (2) in the current implementation of \emph{K2} using only $gri$ imaging, the Balmer break of the elliptical galaxies in the cluster cores, which are the principal contributors to the detection significance, falls outside wavelength range. To address this, we are in the process of incorporating additional ($i-z$) color in available fields and thus improving the completeness at redshifts above $z\,\sim\,0.8$. Finally, the comparison with the MF catalogs have also been consistent with the Monte Carlo estimates. Cases of mismatch arise due to the inherent selection biases present in the fiducial cluster-model based MF method and the color-based $K2$ approach. 

\indent Supported by these successful test results, we have generated $K2$ cluster catalogs for all the $161 \sqdeg$ of CFHTLS-W fields for which \emph{Terapix T05} $g$, $r$ and $i$-band photometric  catalogs are available. Using an automated method, RGB color images of the high likelihood cluster candidates have been generated for a visual search for lensed arc images. We are also improving the photometric redshift values, at present determined using the 2-color method described in the Appendix, to an estimate using \emph{Hyper-Z} \citep{Bolz00} with the full set of $ugriz$ photometry in fields where it is available. The current version of $K2$ cluster catalogs are available on request from the authors.


\acknowledgments

We wish to acknowledge the generous help of David Balam with initial versions of photometric catalogs for the CFHTLS-W fields. The results reported here are based on observations obtained with MegaPrime/MegaCam, a joint project of CFHT and CEA/DAPNIA, at the Canada-France-Hawaii Telescope (CFHT) which is operated by the National Research Council (NRC) of Canada, the Institut National des Science de l'Univers of the Centre National de la Recherche Scientifique (CNRS) of France, and the University of Hawaii. This work is based in part on data products produced at TERAPIX and the Canadian Astronomy Data Centre as part of the Canada-France-Hawaii Telescope Legacy Survey, a collaborative project of NRC and CNRS. This research also used the facilities of the Canadian Astronomy Data Centre operated by the National Research Council of Canada with the support of the Canadian Space Agency. The work reported here forms part of a thesis dissertation during which KT was supported by a UVic Graduate Fellowship and a National Research Council of Canada Graduate Student Scholarship Supplement Program (NRC-GSSSP) Award. 


{\it Facilities:} \facility{CFHT}, \facility{NRC-HIA CADC}, \facility{Terapix}.


\appendix

\section{Photometric redshifts of \emph{K2} cluster candidates} \label{A1photz}
By providing the photometric redshifts (\textit{photo-z}) of the galaxy group and cluster candidates in our $K2$ catalogs, our objective is to increase their utility for other cosmological applications. Using the available two color photometry, we initially attempted to determine the photo-z using available algorithms, e.g., \emph{HyperZ} \citep{Bolz00}, and compared our estimates against the spectroscopic redshifts which were available in the SDSS spectroscopic catalogs for a small sample of our bright ($r \leq 17.7$mag) cluster member galaxies. Though the results were reasonable, our tests also indicated a significant failure rate of $\sim30\%$, mainly due to the uncertainty in determining the appropriate spectral type of the object using just two colors. 

\indent Given this uncertainty due to limited color information, we have adopted a simpler, more direct approach to obtain an estimate of the photo-z of our candidates. Given that $K2$ detects clusters and groups by identifying member galaxies which lie along the \emph{Red Sequence}, as described in Section \ref{ClusSearch}, we make the reasonable assumption that \emph{all} the member galaxies identified by $K2$ are early type E/S0 galaxies. This assumption permits us to determine their photo-z directly using the evolution tracks of early type galaxy colors with redshift (refer to  \S \ref{RevCS} for details and references). Specifically, we build and then use a 2D fitting function to estimate photo-z from the (\emph{g-r}) and (\emph{r-i}) colors of the member galaxies in our catalogs. \citet{Im02} have successfully used a similar approach for photo-z estimation in the Groth strip survey using just the one available $V-I$ color and $I$ magnitude of the galaxies. The \emph{Cluster Red Sequence} (CRS) method \citep{Glad00} estimates redshifts based on this tight color-redshift correlation of cluster early type galaxies. In this Appendix, we describe our method, the calibrations of our fitting function using the public release of photo-z  catalogs for the four CFHTLS-Deep fields\footnote{http://cencos.oamp.fr/} \citep{Ilbe06}  as well as error estimates using spectroscopic redshifts available in the public VIMOS VLT Deep Survey (VVDS) catalogs \citep{LeFe05} for the galaxy sample in the CFHTLS-D1 field (for brevity, we refer to the CFHTLS-Deep photo-z catalogs as \emph{Ilbert} catalogs and the spectroscopic ones as VVDS catalogs). The redshifts in the $Ilbert$ catalogs are based on the five filter $ugriz$ photometry available for the CFHTLS-Deep fields using their \emph{le Phare} photo-z method \citep{Ilbe06}.

\indent We determined that a 2D cubic polynomial is adequate for tracing the evolution of  the (\emph{g-r}) and (\emph{r-i}) colors of early type galaxies with redshift, as was shown by comparing the $\chi^2$ values of the fit with those for other fitting functions. In this fit, we restrict the redshift interval to $z \leq 1$, since the completeness of $K2$ drops off steeply above z = 0.8. The polynomial coefficients are obtained by fitting the available photo-z of \emph{early type} galaxies in the $Ilbert$ catalogs to their (\emph{g-r}) and (\emph{r-i}) colors, using a standard Levenberg-Marquardt least squares minimization technique. We are able  to select this sample of early type galaxies since the galaxies in the $Ilbert$ catalogs are flagged by spectral type, based upon the best fit template during photo-z estimation. In our fit for the coefficients, we only use galaxies where the confidence limit of the photometric redshift determination is listed as 95\% or greater. For the CFHTLS-D1 field, this selection provides 3976 early type galaxies (with $i \leq 24$mag), with a similar number available for the other three fields. Each of the top two panels in Figure \ref{Fig9} plots (\emph{g-r}) and (\emph{r-i}) against the $i$-magnitude of the selected early type galaxies in the CFHTLS-D1; the different colors pertain to galaxies in equal $\Delta z=0.2$ intervals to highlight any redshift dependent trends. The (\emph{r-i})-color shows a more systematic trend with $i$ magnitude compared to the (\emph{g-r}) color, which shows greater scatter mainly due to the degeneracy in this color at redshifts above 0.4 (as seen in the lower left panel). The lower two panels plot the photometric redshifts against each color, to highlight the redshift dependent trend to which we fit our chosen polynomial. The over-plotted points in these panels are the median and 1$\sigma$ scatter in the colors of galaxies binned by $\Delta z=0.05$ to better emphasize the trend with redshift. Also over-plotted as a solid line are the values of our fitting function evaluated at the median values in the bins (projected on to that color - redshift plane), to show the adequacy of the polynomial form for the fitting function.

\indent We use the spectroscopic redshifts in the public VVDS catalog for the CFHTLS-D1 field \citep{LeFe07}\footnote{http://vizier.u-strasbg.fr/viz-bin/VizieR?-source=III/250} to assess the errors in our photo-z determination. By matching ($\alpha, \delta$) positions, we identify our chosen sample of elliptical galaxies from the $Ilbert$ catalog with the corresponding objects in the VVDS catalog; the VVDS catalog does not list their spectral type. Of the matched galaxies, we select only those which have been flagged with spectroscopic redshift confidence exceeding 95\% and which lie below our chosen redshift limit of 1. The selection yields 363 galaxies in common between the two catalogs for the CFHTLS-D1 field. By comparing the spectroscopic and photometric redshifts of the selected early type galaxies, we determine that the median error in the photometric redshifts in the $Ilbert$ catalog is 0.07,  with no redshift dependency. It must be emphasized that the $Ilbert$ photo-z values are determined using a combination of 8 optical and infra-red colors which are available for the CFHTLS-Deep fields (compared to the two colors in the CFHTLS-Wide fields, which we use for cluster detection and for the photo-z estimation). Therefore, the above error value provides a benchmark against which we next compare our redshift estimates as well as the associated errors for the same sample of galaxies .

\indent  Using the (\emph{g-r}) and (\emph{r-i}) colors of the galaxies in the matched sample, we estimate their redshifts using our 2D polynomial. In this estimate, we use the combination of both colors because the slope of the (\emph{g-r}) versus z correlation in the lower redshift range up to $z\sim0.4$ is steeper than that of the (\emph{r-i}) correlation (\emph{lower panels}, Figure \ref{Fig9}; on the other hand, the (\emph{r-i}) slope is steep at redshifts above 0.4 where the (\emph{g-r}) color becomes degenerate. Therefore, using the colors in combination provides improved sensitivity up to $z\sim0.8$, the redshift interval which our cluster candidates span. Figure \ref{Fig10} shows the correlation between our photo-z values and the corresponding spectroscopic redshifts from the VVDS catalog, (\emph{left panel}) and the differences between the two estimates (\emph{right panel}); over-plotted are also the median and standard deviations of the differences computed for galaxies in 0.2 redshift bins. For the sample of early type galaxies, the \emph{median} error in our estimates is $\sim0.1$, which is comparable to the median error in the $Ilbert$ photo-z values for the same sample. However, given that these errors are estimated for a \emph{known set of early type galaxies}, we take this error estimate of our method to be the \emph{lower} limit since we do not take into account errors introduced by the uncertainty in the galaxy type and spectral template mismatch during photo-z determination. 

\begin{figure}  
\epsscale{.80}
\plotone{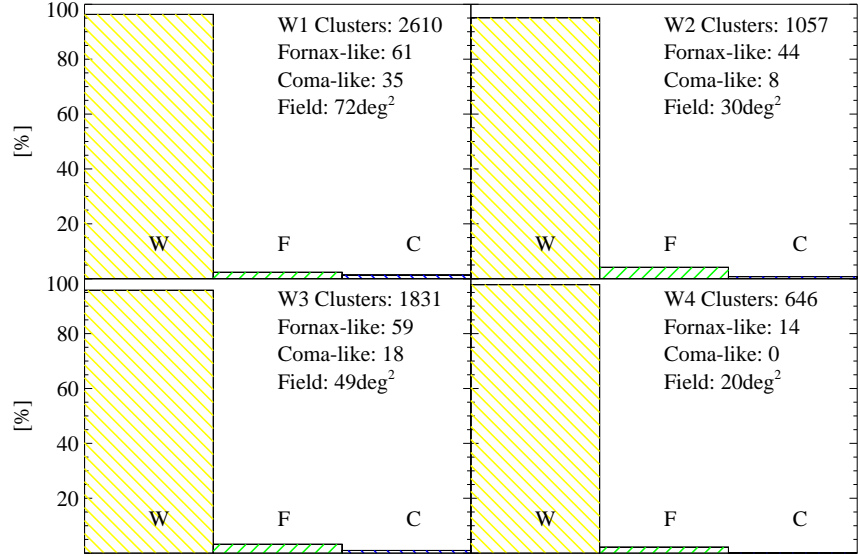}
\caption{Histograms showing the relative percentages of Abell Richness values of the cluster candidates classified as \emph{WBL} (W), \emph{Fornax} (F) and \emph{Coma}-like (C) clusters, corresponding to richness values -1, 0, 1 and higher, respectively. Annotated for comparison are the total number of candidates and areal coverage of each field along with the number of Fornax and Coma-like rich candidates, which have a higher likelihood of lensing background galaxies.\label{Fig8}}
\end{figure}

\begin{figure}  
\epsscale{.80}
\plotone{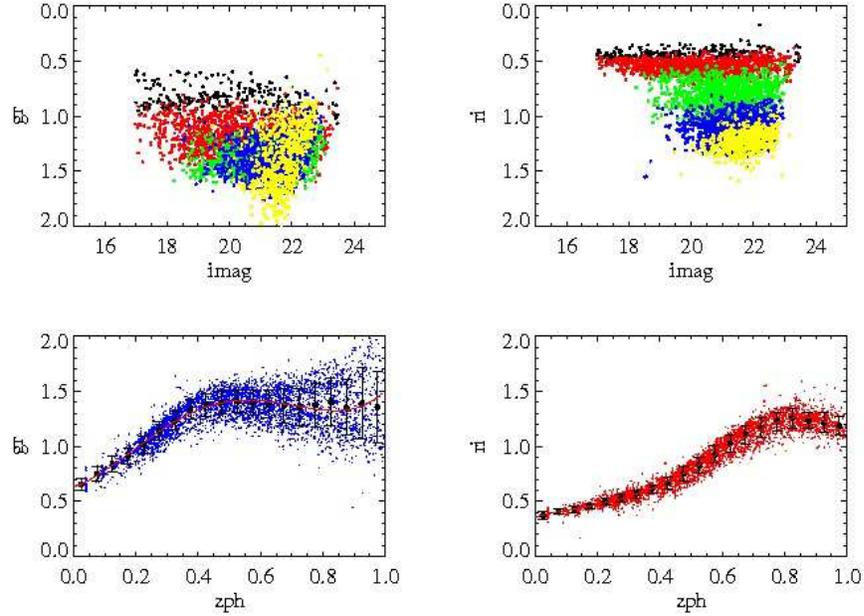}
\caption{CMDs and correlations between photometric redshifts and colors of galaxies fitted with Elliptical galaxy templates for the CFHTLS-D1 field. The top panels show the CMDs for these galaxies, with colors representing redshift bins. The bottom panels are color versus photometric redshifts; the median and standard deviations of colors binned by 0.05 in redshift are over-plotted along with the projected fitted curve for each color (see Appendix \ref{A1photz} for details). All photometric and redshift values are taken from the public release of the photometric redshift catalogs by \citet{Ilbe06}. \label{Fig9}}
\end{figure}

\begin{figure}  
\epsscale{.80}
\plotone{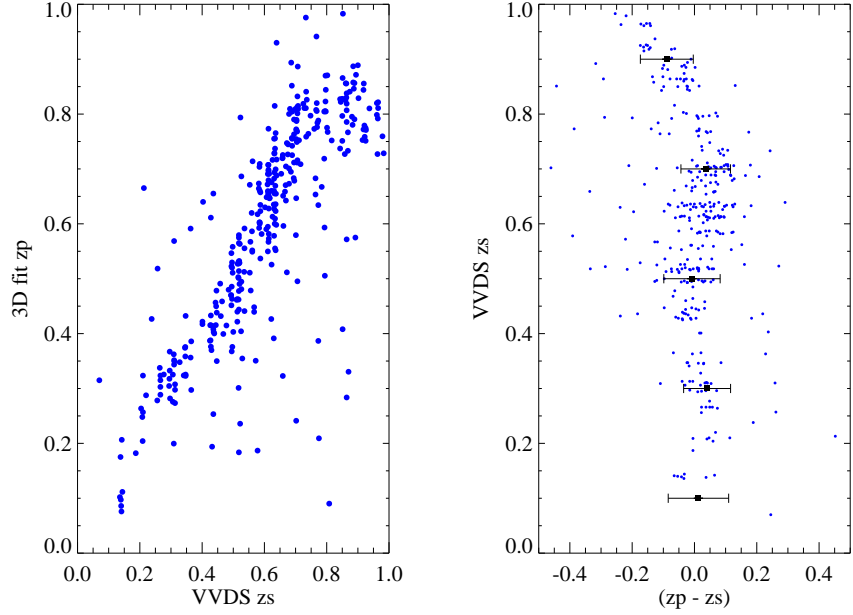}
\caption{ Comparison of the redshifts obtained from our 3D fits with corresponding spectroscopic redshifts from the VVDS catalogs (\emph{left panel}). The sample of 363 early type galaxies selected in the CFHTLS-D1 field is used for this comparison. The \emph{right panel} shows the error in the fit ( = fit redshift  - spectroscopic value) for the same set of galaxies; in order to check for any redshift dependency in this error, the median and $1\sigma$ deviation values, estimated for galaxies in redshift bins of 0.2, are over-plotted. \label{Fig10}}
\end{figure}

\indent In applying our photo-z method to the cluster candidates in the $K2$ catalogs, we obtain the polynomial coefficients for each CFHTLS-Deep field independently to avoid any bias introduced due to color mismatch, e.g. from differing extinction values. We assume that these fits apply to the corresponding CFHTLS-W fields due to the overlap in sky coverage between the Deep and Wide surveys (i.e., the D1 fitting function is used for all the W1 fields, and similarly for the other fields). During the $K2$ cluster detection process, if a BG is flagged as a candidate cluster member, we obtain the photo-z for \emph{all} the galaxies in the detection aperture which also match the BG colors (refer to \S \ref{ClusSearch} for details of the detection). The weighted mean and standard deviation of this set (with the $i$-band magnitudes as weights) are taken to be the photo-z of the BG and the associated error, thus improving the statistical confidence of our photo-z estimate (we also tested the alternative approach of sigma clipping to improve the photo-z estimate; however, as shown in Figure \ref{Fig8}, a significant fraction of the detections are $WBL$-like poor clusters and groups with too few members to make this statistical approach feasible). The estimated photo-z value of the BG is written to the $K2$ catalogs along with the other properties. For single BG cluster candidates, this value is also taken to be the redshift of the cluster. For multi-BG candidates, we take the $i$-magnitude weighted mean and standard deviation of the photo-z of the member BGs to be the redshift of the host group or cluster.     

\clearpage

\bibliographystyle{apj}
\bibliography{ClsFinder}
 
\clearpage




\clearpage

\clearpage

\clearpage

\clearpage

\clearpage

\begin{table}
\begin{center}
\caption{Completion statistics of CFHTLS-W survey \label{Tbl-1}} 
 \scriptsize
 \begin{tabular}{|c||cccccc||cccccc|}
 \tableline
 \tableline
  & \multicolumn{6}{c||}{\textbf{W1}} &  \multicolumn{6}{|c|}{\textbf{W2}}  \\
 \tableline
Filter & N  & A    & I & PSF   & $C_s$ & $C_g$    & N  & A    & I & PSF   & $C_s$ & $C_g$   \\
  &  & [$\sqdeg$]  & [$s$] & [$\asec$]  & [$mag$] & [$mag$] & & [$\sqdeg$]  & [$s$] & [$\asec$]  & [$mag$] & [$mag$] \\
 \tableline
u & 49 & 43.3 & 3000 & 0.91 & 25.38 & 24.7  & 25 & 20.32 & 3000 & 0.91 & 25.41 & 24.7 \\
\textbf{g} & 72 & 63.65 & 2500 & 0.85 & 25.42 & 24.71  & 30 & 24.3 & 2500 & 0.85 & 25.45 & 24.78 \\
\textbf{r} & 72 & 63.75 & 1000 & 0.77 & 24.61 & 23.8  & 32 & 25.87 & 1000 & 0.78 & 24.87 & 24.08 \\
\textbf{i} & 73 & 64.55 & 4300 & 0.71 & 24.45 & 23.68  & 32 & 25.87 & 4300 & 0.67 & 24.75 & 23.73 \\
z & 49 & 43.32 & 3600 & 0.75 & 23.6 & 22.87  & 25 & 20.32 & 3600 & 0.76 & 23.6 & 22.88 \\
 \tableline
 \tableline
  & \multicolumn{6}{c||}{\textbf{W3}} &  \multicolumn{6}{|c|}{\textbf{W4}}  \\
 \tableline
Filter & N  & A    & I & PSF   & $C_s$ & $C_g$  & N  & A    & I & PSF   & $C_s$ & $C_g$   \\
  &  & [$\sqdeg$]  & [$s$] & [$\asec$]  & [$mag$] & [$mag$] & & [$\sqdeg$]  & [$s$] & [$\asec$]  & [$mag$] & [$mag$] \\
 \tableline
u & 30 & 25.57 & 3000 & 0.88 & 25.26 & 24.55  & 25 & 20.87 & 3000 & 0.86 & 25.34 & 24.65 \\
\textbf{g} & 49 & 42.03 & 2500 & 0.89 & 25.43 & 24.73  & 25 & 20.87 & 2500 & 0.8 & 25.42 & 24.75 \\
\textbf{r} & 49 & 42.03 & 1000 & 0.81 & 24.66 & 23.87  & 25 & 20.87 & 1000 & 0.69 & 24.49 & 23.7 \\
\textbf{i} & 49 & 42.03 & 4300 & 0.77 & 24.38 & 23.65  & 20 & 16.86 & 4300 & 0.66 & 24.48 & 23.71 \\
z & 48 & 41.17 & 3600 & 0.73 & 23.57 & 22.85  & 25 & 20.87 & 3600 & 0.68 & 23.5 & 22.8 \\
 \tableline
\end{tabular}
\end{center}
{\footnotesize Completion statistics for the four CFHTLS-W fields computed using information provided at the \emph{Terapix} website, \emph{http://terapix.iap.fr/}. The columns listed are: N = number of fields observed; A = actual sky area covered, accounting for image boundaries; I = integration time; PSF = median seeing during observations; $C_s$ = Completeness limit for stellar objects; $C_g$ = Completeness limit for galaxies. The statistics quoted are the median values for each survey region in the corresponding filter. The three filters used for the current version of the $K2$ cluster catalogs are highlighted in bold font.}

\end{table}
\normalsize 

\clearpage

\begin{deluxetable}{ccccccccc}
\tablecolumns{9}
\tablewidth{0pc}
\tabletypesize{\small}
\tablecaption{Comparison of cluster detections in CFHTLS-Deep fields by the Matched Filter, (MF) and \emph{K2} methods  \label{Tbl-2}}
\tablehead{
\colhead{Field} & \colhead{$K2_{gold}$} & \colhead{$K2_{silver}$} & \colhead{$K2_{bronze}$} & \colhead{$K2_{total}$} & \colhead{$MF_{A}$} & \colhead{$MF_{B}$} & \colhead{$MF_{C}$} & \colhead{$MF_{total}$} }
\startdata
Deep 1 & 18 &  8  &   7  & \textbf{33}  & 19 & 13 & 14 & \textbf{48}    \\
Deep 2 & 29 & 17 & 17 & \textbf{63}   & 17 &  18 & 10& \textbf{45}    \\
Deep 3 & 23 & 19 &  8  & \textbf{50}   &  8& 6 & 6 & \textbf{20}    \\
Deep 4 & 10 & 7   &   6 & \textbf{23}   & 11  &  16  & 9 &\textbf{36}    \\
\enddata
\tablecomments{ Comparison of the numbers of clusters detected in the four CFHTLS-Deep fields by \emph{K2} with those in the Matched Filter Catalog \citep{Olse07}; for each detection method, the breakdown by detection classification as well as the total number are provided (see \S \ref{ClusSearch} and \S \ref{CCDeep} for details of the classification schemes used by each method)} 
\end{deluxetable}


\begin{deluxetable}{cccccc}
\tablecolumns{6}
\tablewidth{0pc}
\tabletypesize{\small}
\tablecaption{MF clusters in CFHTLS-D1 also detected by \emph{K2}  \label{Tbl-3}}
\tablehead{
\colhead{MF class} & \colhead{Number} & \colhead{$K2_{gold}$} & \colhead{$K2_{silver}$}& \colhead{$K2_{bronze}$} & \colhead{$K2_{ND}$} }
\startdata
A class & 19 &  12  &   -  & - & 7  \\
B class & 13 & 2 & 3 & 2  & 6  \\
C class & 14 & 6 &  1  & 1  &  6 \\
\enddata
\tablecomments{ Breakdown of the MF cluster detections in CFHTLS-D1 field, which are also detected by our cluster detection method, \emph{K2}, classified by the assigned class and as non-detections (ND); details of the classification scheme are described in \S \ref{ClusSearch}} 
\end{deluxetable}


\begin{deluxetable}{lcccccccc}
\tablecolumns{9}
\tablewidth{0pc}
\tabletypesize{\small}
\tablecaption{Comparison of \emph{K2} with XMM-LSS X-ray cluster detections in CFHTLS-D1 \label{Tbl-4}}
\tablehead{
\colhead{XMM ID} & \colhead{R.A.} & \colhead{Dec.} & \colhead{$z_{sp}$} & \colhead{K2} & \colhead{$S_{gr}$} & \colhead{$S_{ri}$}  & \colhead{Class} & \colhead{MF} }
\startdata
XLSSC029   & 36.0172   & -4.2247  & 1.05   & N    & 0.00   &  0.00   & - & N \\
XLSSC044   & 36.1410   & -4.2376  & 0.26   & Y   & 12.90    & 9.38   & G& Y \\
XLSSJ022522.7-042648   & 36.3454   & -4.4468 &  0.46  &  Y   & 20.52   &  9.76  & G & N \\
XLSSC025   & 36.3526   & -4.6791  & 0.26  &  Y   &  9.27   & 10.77  & G & Y \\
XLSSJ022529.6-042547   & 36.3733  &  -4.4297  & 0.92   & Y   & 5.27   &  9.23   & G & N \\ 
XLSSC041   & 36.3777   & -4.2388  & 0.14   & Y   & 10.47   & 10.58   & G & Y \\
XLSSC011   & 36.5403  &  -4.9684  & 0.05   & Y   & 11.08    & 3.25   & S & Y \\
XLSSJ022609.9-043120   & 36.5421   & -4.5226  & 0.82   & N    & 2.51    & 7.57  & - & N \\
XLSSC017   & 36.6174   & -4.9967 &  0.38  &  Y   & 15.38   & 10.13   & G & Y \\
XLSSC014   & 36.6411   & -4.0633  & 0.34   & Y   & 14.27    & 7.50   & G & Y \\
XLSSJ022651.8-040956   & 36.7164   & -4.1661  & 0.34   & Y    & 5.03    & 3.37 & S & N \\   
XLSSC005   & 36.7877   & -4.3002  & 1.05   & N   & -0.56   & -0.10   & - & Y \\
XLSSC038   & 36.8536   & -4.1920  & 0.58   & Y    & 3.26    & 3.04  & B & Y \\
XLSSC013   & 36.8588  &  -4.5380  & 0.31   & Y   & 12.53    & 9.66  & G & Y \\
XLSSC022   & 36.9178   & -4.8586  & 0.29   & Y   & 14.39   & 11.64 & G & N \\
XLSSJ022534.2-042535   & 36.3925   & -4.4264 &  0.92   & N    & 1.01    & 3.30  & - & Y \\
XLSSC005b   & 36.8000   & -4.2306  & 1.00   & N    & 1.07    & 1.25  & - & N \\
\enddata
\tablecomments{ Comparison of the detections of the spectroscopically confirmed XMM-LSS X-ray clusters in the CFHTLS-D1 field \citep{Pier07}, using our detection method (\emph{K2}) against the results with the Matched Filter scheme (MF) for the same clusters given in \citet{Olse07}. Our detections are classified further as Gold (G), Silver (S) or Bronze (B), as described in \S \ref{ClusSearch} } 
\end{deluxetable}

\clearpage

\begin{deluxetable}{lcccccccc}
\tablecolumns{9}
\tablewidth{0pc}
\tabletypesize{\small}
\tablecaption{Comparison of Adaptive Kernel \citep{Mazu07} and \emph{K2} cluster detections in CFHTLS-D1 \label{Tbl-4a}}
\tablehead{
\colhead{AK ID} & \colhead{R.A.} & \colhead{Dec.} & \colhead{$z_{c}$} & \colhead{K2} & \colhead{$S_{gr}$} & \colhead{$S_{ri}$}  & \colhead{Class}  & \colhead{d} }
\startdata
	1  &	36.3789   &	-4.2424  &  0.138   &	 Y & 7.33   &  5.97  & G &    91.84 \\
	2   &	36.7981   &	-4.1970  & 0.185   &	 Y &	9.37  &   5.48   & G &   95.46  \\
	3   &	36.3746   &	-4.6831  & 0.225   &	 Y &	 5.00  &   3.43   & S &   75.55 \\
	5   &	36.6240   &	-4.2523  &  0.210  &	 Y &  7.13    &   5.11   & G &  81.21 \\
	10  &	 36.3166  & 	-4.7515   &0.311   &	 Y &	13.15  &   9.33   & G & 118.44 \\
	6   &	36.8416   &	-4.5810   & 0.308  &	 Y &  7.61   &  4.55   & S &   51.65 \\
	12   & 36.6104  & 	-4.5286  & 0.313  &	 Y &  5.32  &   4.69   & S &  120.24 \\
	15   & 36.3842  & 	-4.2726  &  0.542  &	 Y &	11.08  &   5.88  & G &  37.46  \\
	16   & 36.8975  & 	-4.3768   & 0.53  &	 Y &	5.71  &   4.90   & S & 119.98  \\
	17   & 36.8481   &	-4.6202   & 0.543   &	 Y &	7.44  &   5.36   & G &  101.74  \\
	21  &	  36.4645   &	-4.4997  &  0.613  &	 Y & 5.97  &   3.07   & S &   97.78 \\
	19   & 36.8646  & 	-4.5484  &  0.610  &	 Y & 7.75   &  5.84   & G &   66.33 \\
	22   & 36.6686   &	-4.5096  &  0.634  &	 Y & 4.28  &   6.78   & S &  126.25 \\
	28   & 36.8791   &	-4.2025  &  0.784   &	 Y & 4.27   &  7.11   & S &  94.77 \\
 	29   & 36.5030  & 	-4.47142   &  -	  & N & -  & -   & - & - \\
	34   & 36.3925   &	-4.4135  & 0.920  & Y & 3.65   &  4.42    &  B &  90.72  \\

\enddata
\tablecomments{Comparison of the high likelihood cluster candidates in the \emph{AK} catalogs with \emph{K2} detections in the CFHTLS-D1 field. The first four columns list the ID number, the central position in RA and Dec and the redshift of the \emph{AK} detections; these details are reproduced from Table 6 in \citet{Mazu07}. The last five columns list whether \emph{K2} detects the candidate, and if so, the significance values in $gr$ and $ri$ colors, the class of the candidate cluster and the positional offset (in arc sec) between the BCG identified by \emph{K2} and the position given in the $AK$ catalog.} 
\end{deluxetable}


\begin{deluxetable}{cccccccccccccc}
\tablecolumns{14}
\tablewidth{0pc}
\setlength{\tabcolsep}{0.03in}
\tabletypesize{\small}
\tablecaption{Format of master cluster catalog for each CFHTLS-W field \label{Tbl-5}}
\tablehead{
\colhead{ID}  & \colhead{$\alpha$}  & \colhead{$\delta$}  & \colhead{$z_{fit}$}  & \colhead{$i$}  & \colhead{$g-r$}  & \colhead{$r-i$}  & \colhead{$S_{gr}$}  & \colhead{$S_{ri}$}  &  \colhead{Class}   & \colhead{$n_{cls}$}   & \colhead{$n_{32}$}   & \colhead{$\Lambda_r$}   & \colhead{$m_{32}$} }
\startdata
J140252+542504  &  14:02:52.21  &  +54:25:04.77   &   0.279  &  16.61  &   0.887  &   0.486   &  4.64   &  7.44   & 1   & 86  &  83   & 2   & 19.27      \\
J140216+542445 &  14:02:16.96 &  +54:24:45.29  &   0.221 &   17.27 &   0.736 &   0.442 &   4.95 &   7.97   & 1   & 18  &  16  & -1    & 20.26   \\
J140356+542727  & 14:03:56.55  & +54:27:27.26  & 0.314  &  16.69  &  0.342   & 0.238  & 11.13  & 12.22   & 1   & 36  &  33  &  0   & 19.35   \\
J140353+542728  & 14:03:53.60  & +54:27:28.88   &  0.257   & 17.10   & 0.892   & 0.461  &  9.00  & 11.89   &  1  &  21  &  29  & -1  &  20.52    \\
J140646+542503 &  14:06:46.01 &  +54:25:03.16  &   0.352  &  17.51  &  1.399 &   0.614  &  6.59 &   5.10   & 1   &  3   &  7  & -1    & 21.27   \\
\enddata
\tablecomments{ A sample listing of a typical $K2$ cluster catalog generated for each 1 $\sqdeg$ CFHTLS-W field. The properties listed are: a unique ID number, J2000 sky positions ($\alpha$ and $\delta$), a photometric redshift estimate, $i$ magnitude and (\emph{g-r}) and (\emph{r-i}) colors of the BCG, the (\emph{g-r}) and (\emph{r-i}) detection significances, the \emph{K2} detection class, the number of bright cluster members, the number of cluster members included in the richness class, the corresponding Abell richness and the limiting magnitude, $m_{32}$, used in computing the Abell richness. The positional and photometric details are taken directly from the $Terapix$ catalogs; all the other properties are estimated during the detection process. \emph{Complete cluster catalogs for the $161\sqdeg$ CFHTLS-W fields are available on request from the authors}}. 
\end{deluxetable}

\clearpage

\begin{deluxetable}{ccccccccccc}
\tablecolumns{11}
\tablewidth{0pc}
\setlength{\tabcolsep}{0.05in}
\tablecaption{Format of supplementary catalog for cluster member properties for each candidate cluster \label{Tbl-6}}
\tablehead{
\colhead{ID} & \colhead{Class} & \colhead{$i$} & \colhead{$g-r$} & \colhead{$r-i$} & \colhead{$z_{fit}$}  & \colhead{$x$}  & \colhead{$y$}  & \colhead{$HLR$} & \colhead{$S_{gr}$} & \colhead{$S_{ri}$} }
\startdata
 137343   & 1  &  16.607   &  0.887  &   0.486   & 0.271 & 16837.26  &  7955.87   & 10.64  &   12.64   &  7.44   \\
 133713  &  1   & 17.444   &  0.914   &  0.565   & 0.294 & 16893.84   & 7718.00   &  7.53   &  4.42    & 4.65  \\
 153526   & 1   & 19.021  &   0.719   &  0.568   & 0.269 & 16918.31   &  8337.70   &  7.09   &  9.22   &  8.45  \\
 153427  &  1   & 19.038   &  0.897   &  0.565  & 0.287 & 16956.09  &  8362.39   &  5.90   &  7.25   &  7.59  \\
\enddata
\tablecomments{ A sample listing of the supplementary catalog generated by $K2$ for any cluster or group in which multiple BGs have been detected. The columns listed are a unique ID number for each BG, $K2$ class of the cluster, $i$ magnitude and (\emph{g-r}) and (\emph{r-i}) colors, photometric redshift, pixel positions in the corresponding $Megaprime$ image and the (\emph{g-r}) and (\emph{r-i}) detection significances. \emph{Complete set of supplementary catalogs are available along with the corresponding master catalogs on request from the authors}}. 
\end{deluxetable}


\begin{deluxetable}{lllll}
\tablecolumns{5}
\tablewidth{0pc}
\setlength{\tabcolsep}{0.05in}
\tablecaption{Detection statistics in $K2$ catalogs for each CFHTLS-W field \label{Tbl-7}}
\tablehead{
\colhead{Statistic} & \colhead{W1} & \colhead{W2} & \colhead{W3} & \colhead{W4} }
\startdata
Number of fields & 72 & 30 & 49 & 20 \\
Total detections & 2610 & 1057 & 1831 & 646 \\
Detections/$\sqdeg$ & $35\pm12$ & $35\pm9$  & $37\pm15$  & $30\pm14$  \\
\cline{1-5}  \\
Class & & & & \\
\cline{1-5}  \\
Gold & 984 & 357 & 713 & 251 \\
Silver & 737 & 301  & 520 & 163 \\
Bronze & 889 & 399 & 598 & 232 \\
Gold [\%] & $37\pm16$  & $34\pm13$ & $38\pm20$  & $40\pm16$ \\
Silver [\%]  & $28\pm16$ & $28\pm16$ & $28\pm19$ & $23\pm17$ \\
Bronze [\%]  & $34\pm18$ & $39\pm16$  & $34\pm17$ & $37\pm15$  \\
\cline{1-5}  \\
Richness & & & & \\
\cline{1-5}  \\
WBL total [\%] & 2514 [96.3] & 1005 [95.1]  & 1754 [95.8] & 632 [97.8] \\
 Fornax total [\%]  & 61 [2.3] & 44 [4.2]  & 59 [3.2] & 14 [2.2] \\
 Coma total [\%]  & 35 [1.3]  & 8[0.8]  & 18 [1]  & 0 [-] \\
WBL [number/$\sqdeg$]  & 34.9 & 33.5 & 35.796 & 31.6 \\
Fornax  [number/$\sqdeg$]   & 0.847 & 1,467 & 1.204 & 0.7 \\
 Coma [number/$\sqdeg$]    & 0.486 & 0.267 & 0.367 & - \\
 WBL [$\sqdeg$/cluster] & 0.029 & 0.03 & 0.028 & 0.032 \\
Fornax [$\sqdeg$/cluster]  & 1.18 & 0.68 & 0.83  & 1.43 \\
Coma [$\sqdeg$/cluster]   & 2.06  & 3.75 & 2.72 & -  \\
\enddata
\tablecomments{ Comparison of the detection statistics for the four CFHTLS-W fields deduced from the $K2$ cluster catalogs. The details listed are: the total number of fields (= number of fields with $ Terapix\; T05\;  g, \; r$ and $i$ photometric catalogs), total number of detections, number of detections/$\sqdeg$, detections in each detection class/$\sqdeg$, total number of detections in each Abell richness class, the surface density of each Abell richness class, and the inverse surface density (= sky area in which 1 cluster of that Abell richness class is found). All the quoted statistics are corresponding median and inter-quartile distances. The percentages listed in square brackets for the Abell richness classes (Rows 10 - 12) are computed using the total number of detections in that survey region, given on Row 2 of this table.} 
\end{deluxetable}
\clearpage


\end{document}